\documentclass[12pt]{article}

\usepackage{algorithm,algpseudocode}
\usepackage{amscd,amsfonts,amsopn,amssymb,amstext}
\usepackage{appendix}
\usepackage{booktabs}
\usepackage{bbm,bm}
\usepackage[centertags]{amsmath}
\usepackage{color}
\usepackage{fullpage}
\usepackage{graphicx,graphics,psfrag}
\usepackage{indentfirst}
\usepackage{latexsym,enumerate}
\usepackage{lscape}
\usepackage{multirow}
\usepackage{natbib}
\usepackage{rotating}
\usepackage{setspace}
\usepackage{srcltx}
\usepackage{subfigure}
\usepackage[T1]{fontenc}
\usepackage{threeparttable}
\usepackage{times}
\usepackage{url}
\usepackage{verbatim}

\usepackage[colorlinks=true, linkcolor=blue, citecolor=blue, urlcolor=blue]{hyperref}
\usepackage{natbib}

\makeatletter
\renewcommand\@biblabel[1]{#1.}
\makeatother

\newcommand{\qed}{\rule{0.5em}{1.5ex}}

\title{\Large \bf{A two-stage model for factors influencing citation counts}}

\author{\normalsize
{\bf Pablo Dorta-Gonz\'alez\thanks{P. Dorta-Gonz\'alez, Department of Quantitative Methods in Economics,
University of Las Palmas de Gran Canaria, 35017 Las Palmas de Gran Canaria, Spain. E-mail: \texttt{pablo.dorta@ulpgc.es}}, Emilio G\'omez-D\'eniz }\\
{\small Department of Quantitative Methods in Economics and TIDES Institute,}\\[-0.2cm]
{\small University of Las Palmas de Gran Canaria, Spain}\\[-0.15cm]
{\small http://orcid.org/0000-0003-0494-2903(PDG) }\\[-0.20cm]
{\small http://orcid.org/0000-0002-5072-7908 (EGD)}
}

\date{}

\def \E{{\rm I\kern -2.2pt  E}}

\begin{document}

\maketitle

\vspace{-1cm}

\begin{abstract}\noindent
This work aims to study a count response random variable, the number of citations of a research paper, affected by some explanatory variables through a suitable regression model. Due to the fact that the count variable exhibits substantial variation since the sample variance is larger than the sample mean, the classical Poisson regression model seems not to be appropriate. We concentrate attention on the negative binomial regression model, which allows the variance of each measurement to be a function of its predicted value. Nevertheless, the process of citations of papers may be divided into two parts. In the first stage, the paper has no citations, and the second part provides the intensity of the citations. A hurdle model for separating the documents with citations and those without citations is considered. The dataset for the empirical application consisted of 43,190 research papers in the field of Economics and Business from 2014-2021, obtained from The Lens database. Citation counts and social attention scores for each article were gathered from Altmetric database. The main findings indicate that both collaboration and funding have a positive impact on citation counts and reduce the likelihood of receiving zero citations. Higher journal impact factors lead to higher citation counts, while lower peer review ratings lead to fewer citations and a higher probability of zero citations. Mentions in news, blogs, and social media have varying but generally limited effects on citation counts. Open access via repositories (green OA) correlates with higher citation counts and a lower probability of zero citations. In contrast, OA via the publisher's website without an explicit open license (bronze OA) is associated with higher citation counts but also with a higher probability of zero citations. In addition, open access in subscription-based journals (hybrid OA) increases citation counts, although the effect is modest.

\vspace{-0.5cm}
\vspace{-0.5cm}
\vspace{-0.5cm}
\end{abstract}
\vspace{-0.5cm}

\section{Introduction}
Count data regression models arise in situations in which the variable of interest takes only nonnegative integer
values for each of the available observations. These values usually represent the number of times an event occurs in a fixed domain. \cite{cameronandtrivedi_1998} and \cite{winkelmann_2008} provide good overviews of standard count regression models. Typically, a Poisson distribution could be assumed to model the distribution of the count citations. Nevertheless, the Poisson model often underestimates the observed overdispersion (variance larger than the mean). This is so because a single parameter probably will be insufficient to describe the population under study, and since this population is usually heterogeneous. This population heterogeneity is
unobserved, i.e., the population consists of several subpopulations, and the general method to deal with this is to assume that the heterogeneity involved  can be adequately described
by some probability density function, say $\pi$, defined on the population of possible Poisson parameters, $\lambda>0$, and consider the marginal distribution of the number of citations given by
\begin{eqnarray}
f(y)=\int_0^{\infty}\exp(-\lambda)\frac{\lambda^y}{y!} \pi(\lambda)\,d\lambda,\quad y=0,1,\dots\label{mp}
\end{eqnarray}

Mixed Poisson distributions are helpful in situations where counts display
extra-Poisson variation. Applications of univariate models abound in areas such as insurance
\citetext{\citealp{willmot_1987}, and \citealp{gomezandcalderin_2018}} and accident analysis \citetext{\citealp{arbousandkerrich_1951}}, where specific models such
as the negative-binomial (mixed Poisson-gamma), Poisson-inverse-Gaussian, Poisson-reciprocal inverse-Gaussian and Poisson-lognormal distributions have been used. On the other hand, mixed Poisson
regression models have been employed in areas such as insurance \citetext{\citealp{deanetal_1989}}, demography \citetext{\citealp{brillinger_1986}}, medicine \citetext{\citealp{breslow_1984}, and \citealp{campbelletal_1991}}, and engineering \citetext{\citealp{engel_1984}}. For a complete revision about mixed Poisson distributions, see \cite{karlisandxekalaki_2005}, and for Poisson and negative binomial regression models, see \cite{cameronandtrivedi_1998}, \cite{winkelmann_2008} and \cite{hilbe_2011}, among others.

Nevertheless, the process of citations of papers may be divided into two parts. In the first stage, the paper has no citations, and the second part provides the intensity of the citations. Once a paper is cited for the first time, the contagious process of getting new citations is carried out, at most for the time following this first citation. A hurdle model for dealing separately with the papers with citations and the papers with no citations is then considered.

{The hurdle model in bibliometrics and altmetrics represents a significant advance in understanding the citation process. Several studies and experts support this idea. For example, studies in biology, biochemistry, chemistry, and social sciences \citetext{\citealp{didegahandthelwall_2013}}, as well as a study of permanent Italian researchers from different scientific fields \citetext{\citealp{baccini_2014}}, found that the citation process can be modeled using a hurdle approach. In this model, the initial publication phase is characterized by a low probability of citation, while the subsequent phase has a higher probability.

Furthermore, the hurdle model has important implications for the assessment of research impact. It suggests that the initial publication phase is not necessarily indicative of the overall impact of a research paper. Instead, the critical point at which citations begin can strongly influence the total number of citations. This is particularly relevant in the context of altmetrics, where initial engagement with a research paper can trigger subsequent mentions and other forms of online engagement \citetext{\citealp{hodas_2014}}.
}

It is important to note that human behavior when deciding which works to cite is complex. The aim of this study is not to predict such behavior or to estimate citation counts, but rather to contribute knowledge regarding some of the variables involved. Over the past two decades, there has been a paradigm shift in how scientists disseminate their research findings. On one hand, they tend to publish in journals with high impact factor rankings; on the other, they increasingly favor open access and dissemination through channels that are closer to society, in addition to traditional academic outlets. The goal of this study is to provide insights into the effects of these new forms of scientific communication on research impact.

The remainder of this paper is organized as follows. {Section \ref{s1} focuses on reviewing the literature.} Section \ref{s2} describes the data set and variables used in this work. Section \ref{s3} is devoted to the specific distributional models employed to reach this work's target. Numerical results are provided in Section \ref{s4}  and conclusions in the last Section \ref{s5}.

{
\section{Literature revision: Factors linked to more citations or better quality\label{s1}}
Evaluation of research is commonly based on citation metrics and expert evaluations. However, new research indicates that incorporating additional variables can enhance prediction accuracy. While previous research emphasized citations as a measure of quality, not many studies have examined expert reviews, especially in areas such as humanities and some social sciences where quality may not be accurately represented by citations. This literature review aims to examine how consistent factors influencing document impact are across various contexts and to determine when these factors are dependent on the context.

\subsection{Internal factors influencing citation counts and quality}

Scientists have thoroughly studied factors associated with increased citation numbers or projected citation counts over time using journal or article metadata. For example, research has indicated that papers with higher peer review ratings are more likely to be cited \citetext{\citealp{bornmann_2012}}. Furthermore, \cite{koushaandthelwall_2024} conduct an extensive examination of different document-level characteristics and approaches employed in forecasting citation counts and quality assessments of scholarly articles.

The majority of research has studied document characteristics by evaluating how they are linked to the number of citations received. For example, a research study looked at various aspects of article texts (such as length of title and text in characters, and number of figures, tables, and equations), metadata (like number of authors and views), and citation numbers from two hundred papers published by MDPI in 2017 which were either highly cited or lowly cited. \cite{elgendi_2019} discovered strong connections between citation counts and the number of views, tables, and authors, and also a noticeable inverse relationship with title length. A study of 262 papers found that there was a weak correlation between "non-scientific features" and citation counts (Pearson's |r| < 0.2), while the relationship between journal impact factors or the number of authors and citation counts grew stronger as time went on \citetext{\citealp{mammola_2022}}. Furthermore, certain research studies have utilized regression models to examine numerous characteristics at the same time, enabling the evaluation of their respective impacts \citetext{\citealp{alpay_2022}}.

Articles with more authors could have better quality because they involve a wider range of expertise or deal with complex research that requires multiple contributors. On the other hand, a greater amount of writers may also spark additional attention via their wide social connections, leading to an audience impact, making it challenging to distinguish between cause and effect \citetext{\citealp{rousseau_1992}; \citealp{wagner_2019}}. Although previous studies have found a connection between the number of citations and the number of authors, there is no agreement on the specific type of relationship, such as linear or logarithmic. Bigger groups of authors frequently involve a greater number of organizations and nations, which can impact the connection between the number of citations and team size. However, research that has taken these factors into consideration still shows that bigger teams have a citation advantage.

Many research studies in different disciplines consistently show that articles with a greater number of authors generally receive more citations. This pattern is commonly seen and strong. Studies in prestigious interdisciplinary journals such as Nature, Science, and PNAS, along with biomedical and scientific journals, have brought attention to this trend \citetext{\citealp{hsuandhuang_2011}}. Likewise, research in respected publications like Cell, Science, Nature, New England Journal of Medicine, The Lancet, and JAMA have verified this connection \citetext{\citealp{figg_2006}}. Similar results have been observed in fields such as biology, biochemistry, chemistry, mathematics, physics \citetext{\citealp{vieira_2010}}, library and information science \citetext{\citealp{sin_2011}}, computer science \citetext{\citealp{ibanez_2013}}, natural and medical sciences, social sciences, humanities \citetext{\citealp{lariviere_2015}}, management \citetext{\citealp{ronda-pupo_2017}}, and robotics and artificial intelligence \citetext{\citealp{kumari_2020}}.

Valuable knowledge has been acquired from research that highlights specific countries or organizations, like Italy \citetext{\citealp{abramoanddangelo_2015}}, as well as Belgium, Israel, and Iran \citetext{\citealp{chiandglanzel_2017}}. A thorough examination across ten countries, focusing on 27 broad subjects and the highest volume of journal articles from 2008 to 2012, uncovered a strong connection between increased collaboration and higher citation rates in most subjects and countries. Nevertheless, in areas such as computer science, business, management, and accounting, China showed weaker correlations between the number of authors and citation counts \citetext{\citealp{thelwall_2020}}. Additionally, the connection between citations and collaboration in research could be stronger in developing countries than in developed ones according to \cite{shen_2021}.

It is widely supported by evidence from different areas that articles in journals with higher impact factors are typically cited more frequently. This pattern has been consistently noted in numerous studies, including research on biology, biochemistry, chemistry, mathematics, and physics articles \citetext{\citealp{vieira_2010}}, papers published in F1000 from 2000 to 2004 \citetext{\citealp{bornmannandleydesdorff_2015}}, and studies focusing on six biomedical research topics from 1990 to 2018 \citetext{\citealp{urlings_2021}}. These studies indicate that the journal impact factor is frequently recognized as the most reliable bibliometric indicator of article citations.

Although many research studies have discovered a strong relationship between journal impact factor and the number of times articles are cited, there are some cases where this is not the norm. Studies in ecology \citetext{\citealp{leimuandkoricheva_2005}} and gastroenterology and hepatology \citetext{\citealp{roldan-valadez_2015}} failed to uncover enough statistical proof to back up the idea that papers in high-impact factor journals receive more citations. The absence of a connection in these instances could be due to limited sample sizes or the distortion caused by a few heavily referenced articles on a journal's overall impact factor.

Experts are often considered the most appropriate evaluators of research quality, with citation counts being, at most, a potential indicator of it. The three main aspects of academic research are usually defined as rigor, originality, and importance to the scholarly and societal community according to \cite{aksnes_2019} and \cite{langfeldt_2020}. Every dimension is subjective and shows substantial differences among various fields.

Different expert evaluations of the quality of academic research can occur because of the diverse viewpoints used to assess quality \citetext{\citealp{langfeldt_2020}}. Additionally, high-quality research in a certain field that is in line with the field's goals may not be recognized at the same level in national research assessments. This difference may arise when the objectives of the field are not clear or are ignored, especially if they are seen as too abstract and do not take into account societal viewpoints.

\subsection{External factors influencing citation counts and quality}

Although various factors have been identified that can predict citation impact or quality scores, the connections between these factors and research quality are frequently dependent on the specific context. The extent to which these relationships are strong differs greatly among various academic disciplines. In the physical sciences, citation-related information is typically more predictive than in the humanities, as shown by \cite{dortaandgomez_2022}. Moreover, the way citations are used and the impact of specific factors may evolve with time, requiring regular adjustments to predictive models to account for these changes over time.

Open Access (OA) articles appear to receive more citations because they are more widely accessible. Nevertheless, validating this benefit is made difficult by the diverse forms of open access and different journal characteristics, with both top-tier and lower-quality journals being capable of being completely open access or completely non-open access \citetext{\citealp{dortagonzalez_2017}; \citealp{dortagonzalez_2023a}}. Furthermore, it is difficult to consider author choices, such as if scholars tend to publish their top research as OA. The intricate nature of these factors, combined with possible variations across disciplines, leads to the uncertainty surrounding the existence of a real advantage in OA \citetext{\citealp{langhamputrow_2021}}.

According to \cite{thelwall_2023}, UK articles that disclose their funding source generally exhibit higher quality in all fields, irrespective of research team size. This phenomenon is especially prominent in areas related to health. The research indicates that funding is important for quality research as it involves scrutiny and validation which unfunded projects may lack.

Research articles in quickly growing research fields, like newly developing trends, usually get more citations than the average for that field. The increase in citations in these fields is due to the rise in publications with limited existing literature to reference, resulting in a more concentrated number of citations \citetext{\citealp{sjogarde_2022}}.

Research has indicated that incorporating both early citation counts and internal/external factors can forecast long-term citation counts successfully. A neural network was used in a study to predict the 5-year citation impact of articles in the library, information, and documentation field. This research included a variety of elements, such as metadata from article text, journals, authors, references, and citations. The characteristics included different factors like the type of document, length of the article, journal impact factor, number of authors, previous citations, and others, leading to favorable outcomes \citetext{\citealp{ruan_2020}}. In another method, a different study utilized both altmetric indicators and metadata to anticipate upcoming citations for a random selection of 12 thousand articles released in 2015. The study found that machine learning models that included variables such as Mendeley readership, highest number of Twitter followers, and academic standing were important indicators of citation impact in both the short and long term \citetext{\citealp{akella_2021}}. This research emphasizes the significance of taking into account various factors when predicting long-term citation numbers.

Scientific research has significance outside of academia, affecting society in fields such as education, culture, and the economy \citetext{\citealp{wilsdon_2015}}. Altmetrics can quantitatively assess the broader societal impact by using a complementary set of metrics known as altmetrics, which is different from traditional citation analysis. The changing landscape of digital academic communication has changed the way we assess the societal influence of research, promoting a broader method that takes into account a variety of research outcomes and new communication methods \citetext{\citealp{bornmann_2013}; \citealp{derijcke_2016}; \citealp{bornmannandhaunschild_2019}}. An example is the UK's Research Excellence Framework, which includes evaluations of research impact outside of academia, with 25\% of the assessment focusing on areas such as impact on public policy, economic and social contributions, and advancements in health, environment, and overall well-being \citetext{\citealp{khazraguiandhudson_2015}}.

Initial altmetrics studies concentrated on internet references and their connection to citations, suggesting uniform social influence for all references, without a strong theoretical foundation \citetext{\citealp{ravenscroft_2017}}. Academic evaluation literature differentiates between scientific impact in academia and wider societal impact \citetext{\citealp{spaapen_2011}; \citealp{joly_2015}}. While altmetrics initially gained traction as a measure of societal impact due to funding agencies' interests, the field now stresses the importance of a more sophisticated approach. Recent research suggests utilizing altmetrics to assess science-society interactions and knowledge sharing, instead of traditional impact metrics \citetext{\citealp{haustein_2016}; \citealp{robinson-garcia_2018}; \citealp{wouters_2019}; \citealp{dortagonzalez_2023b}; \citealp{alperin_2024}; \citealp{dortagonzalez_2024}}.

}

\section{The data\label{s2}}
{
\subsection{Data sources and sample description}

In the empirical application, we explore research trends in Economics and Business by examining scientific articles. This analysis focuses on the publication period 2014-2021 and the citation period 2014-2023, using data from The Lens, Scimago Journal Rank (based on Scopus data), Altmetric, and the Australian Business Deans Council.

We used the Field of Research (FoR) classification provided by The Lens as a classification system for determining the discipline. This system is also used by other bibliographic databases and is generated automatically using artificial intelligence. It is important to note that this AI-based classification has not yet been thoroughly validated, and its accuracy and consistency across different fields of research are uncertain. Nevertheless, we consider it suitable for the purposes of this study.

The analysis focused on journal articles as the document type and was limited to the years 2014 to 2021. However, citation data were collected up to 2023 to include the two years following publication — typically the period during which citation counts peak.
We acknowledge that including publications from this period may introduce a citation lag, particularly for more recent articles. However, our model includes time as an explanatory variable, enabling us to isolate and compare the effects of other variables in relation to the passage of time. This approach enables us to explicitly account for delays in the accumulation of citations, thereby reducing potential bias associated with newer publications.

The search criteria in The Lens were the following. Field of Study: (Business OR Economics); Publication Date: (2014-01-01 TO 2021-12-31); Publication Type: (journal article); Institution Country = (Australia, Brazil, Canada, China, France, Germany, India, Indonesia, Italy, Japan, Netherlands, Republic of Korea, Russia, Spain, United Kingdom, United States).

The analysis focuses on the 16 countries with the highest production of Business and Economics articles over the period analyzed. This selection was made to ensure a representative sample of Economics and Business articles over the last decade. To allow for the accumulation of citations, the search was limited to articles published up to 2021.

The following article-level variables were obtained from The Lens database: publication year, ISSNs, number of authors, funding, DOI, number of citations, open access, OA type.

Next, the ISSNs were searched in the Scimago Journal Rank database to obtain the following variables at the journal level: foundation year (proxy for the year of inception in Scopus), SJR, SJR best quartile, citations per document (3-year period).

Also using the ISSNs, the Australian Business Deans Council journal quality list \citetext{\citealp{abdc_2022}} was linked to obtaining the expert rating and the Fields of Research (FoR) code at the journal level. In 2022, the expert rating process resulted in a total of 2,680 journals receiving classifications, with the distribution as follows, from highest to lowest: A* = 7\% (199), A = 25\% (653), B = 32\% (855), and C = 36\% (973). Every journal on the list must fall within relevant Australia and New Zealand Fields of Research (FoR) codes.

The aggregation into disciplines was carried out according to the following (code and FoR): Accounting and Finance (3501 Accounting, auditing and accountability; 3502 Banking, finance and investment), Applied Economics (3801 Applied economics; 3802 Econometrics), Business (3505 Human resources and industrial relations; 3506 Marketing; 3507 Strategy, management and organisational behaviour), Commercial (4801 Commercial law; 3504 Commercial services; 3599 Other commerce, management, tourism and services), Economic Theory (3803 Economic theory; 3509 Transportation, logistics and supply chains), Statistics (4905 Statistics), Tourism (3508 Tourism). The discipline of Statistics was used as a reference point for comparison within the regressions.

Finally, we queried the Altmetric.com database using the Digital Object Identifier (DOI) of each article to obtain a set of altmetric indicators at paper level. Specifically, data was collected on mentions in the news, on blogs, in policy documents and patents, on X (formerly Twitter) and Facebook, in Wikipedia citations, on video platforms (e.g. YouTube) and the number of Mendeley readers. This information was retrieved using the Altmetric.com search interface, with the DOI serving as the primary search criterion to ensure precise matching between records. Using the DOI guarantees a high level of accuracy in data linkage as it serves as a unique and persistent identifier for each publication. Each altmetric variable was recorded as a raw count, without weighting or composite scoring, to enable disaggregated analysis of the effects of different types of online attention on citations. 

Data were downloaded and merged during the last two weeks of March 2024.

\subsection{Variable description}

Table \ref{variables} describes the variables analyzed and their coding in cases where it was necessary. In terms of access type definitions, closed access, also known as subscription access, refers to the traditional model in which scholarly articles are only available to readers through subscription or paywall barriers. Typically, articles are available only to subscribers or individuals affiliated with subscribing institutions, thereby limiting access to a wider audience. Gold OA, on the other hand, is the practice of publishing scholarly articles in fully open access journals where the articles are freely available to readers without subscription or paywall restrictions. These articles are usually published under a Creative Commons license, which allows them to be freely distributed. Authors may be charged an article processing charge (APC) to cover publication costs.

On the other hand, hybrid OA refers to the publication of individual articles in subscription-based journals, with the option for authors to pay a fee for open access to their articles. This model allows journals to retain subscription revenue while offering authors the choice of open access publication. Green OA involves the self-archiving or deposit of scholarly articles in repositories or platforms after publication in subscription-based journals. These articles become openly accessible through institutional repositories, subject-based repositories, or preprint servers, extending access beyond the journal's paywall. Finally, bronze OA is the practice of making articles openly accessible on a publisher's website without an explicit open license. Some publishers choose to make selected articles freely available within subscription-based journals or designate specific journals or sections where articles are accessible without a subscription. In some cases, publishers may impose an embargo period during which articles remain behind a paywall. After this period, the articles become freely available, a practice known as delayed open access.

}

\begin{table}[htbp]
\resizebox{0.75\textwidth}{!}{\begin{minipage}{\textwidth}
\caption{Description of factors possibly influencing the number of citations}\label{variables}
\begin{center}
\begin{tabular}{ll}
Variable & Description\\ \hline
{\bf Access type} & Closed; Gold; Hybrid; Green; Bronze\\ \\ [-0.30cm]
{\bf Year of publication} & Takes values from 2014 to 2021. Measures the age of the article\\ \\ [-0.30cm]
{\bf Discipline} & Accounting and Finance; Applied Economics; Economic Theory;\\
&  Business; Commercial; Tourism; Statistics\\ \\
{\bf Number of authors} & Takes values from 1 to 47. Measures the collaboration\\ \\ [-0.30cm]
{\bf Funding} & Takes the value 0 if the paper has not been financed and 1 otherwise\\ \\ [-0.30cm]
{\bf Expert rating} & Takes values 4 (top 7\% of journals); 3 (next 25\%); 2 (next 32\%); 1 (bottom 36\%).\\
&   Measures experts' assessment of the journal\\ \\ [-0.30cm]
{\bf Foundation year} & First year of indexing in the Scopus database. A proxy for the foundation year, measures\\
&  prestige by signaling history, credibility, and longevity\\ \\ [-0.30cm]
{\bf SJR} & The Scimago Journal Rank measures the prestige of journals.\\
&   It takes into account both the quantity and quality of citations, with citations\\
&  weighted according to the influence of the citing journal\\ \\ [-0.30cm]
{\bf SJR best quartile} & Takes values from 1 (top) to 4 (bottom). Ranks journals into quartiles\\
&  according to their SJR scores\\ \\ [-0.30cm]
{\bf Cites per document} & Average number of citations per article in the journal over a three-year period\\ \\ [-0.30cm]
{\bf News mentions} & Times an article has been mentioned in news articles.\\
&   Measures the reach and influence beyond the academic sphere\\ \\ [-0.30cm]
{\bf Blog mentions} & Times an article has been referenced or discussed in blog posts.\\
&   Measures the impact on both academic and non-academic online discussions\\ \\ [-0.30cm]
{\bf Policy mentions} & Times an article is cited in policy documents.\\
&   Measures its importance in informing policy decisions and shaping public discourse\\ \\ [-0.30cm]
{\bf Patent mentions} & Frequency with which an article is cited in patents. Measures the relevance and potential\\
&  application to innovation and technological advancement\\ \\ [-0.30cm]
{\bf X mentions} & Times an article was referenced, shared, or discussed on the leading social media\\
& platform, X/Twitter. Measures visibility, impact, and engagement within the\\
&  X/Twitter community\\ \\ [-0.30cm]
{\bf Facebook} & Times an article was mentioned or shared on Facebook.\\
&  Indicates its popularity and visibility on this widely used social media platform\\ \\ [-0.30cm]
{\bf Wikipedia} & The frequency with which an article is referenced on Wikipedia pages.\\
&  Measures its importance and impact in shaping knowledge and information on the web\\ \\ [-0.30cm]
{\bf Video mentions} & The times an article is referenced on YouTube, including\\
&   lectures, presentations, or online educational content.\\
 &  Serves as a measure of its impact beyond traditional written media\\ \\ [-0.30cm]
{\bf Mendeley} & The number of Mendeley users who have added the article to their library.\\
&  Measures the popularity and relevance of the article within the academic community\\ \hline
\end{tabular}
\end{center}
\end{minipage}}
\end{table}

Table \ref{tabattributes} shows the descriptive statistics obtained for the
dependent and explanatory variables associated with the filtered
database. The large sample size of 43,190 observations provides a strong basis for regressions. In terms of the characteristics of the research articles in the sample, the average article has 27.3 citations, with a median of 11, indicating a right-skewed distribution. The number of citations varies considerably, ranging from 0 to 2672. Over the period analyzed, closed access was predominant in Economics and Business, with 55\% of the observations in the sample, compared to 45\% of open access articles. OA green is the most common type of open access (30\%), followed by OA hybrid (10\%), OA bronze (3\%), and OA gold (2\%).

\begin{table}[h!]
\resizebox{0.80\textwidth}{!}{\begin{minipage}{\textwidth}
\caption{Statistics of the variables}\label{tabattributes}
\begin{center}
\begin{tabular}{lrrrrrr}\hline
Variable & Mean & Median & SD & Min & Max & Relative frequency of 1\\
&&&&&& (dichotomous variables)\\ \hline
\multicolumn{7}{l}{\bf Variables associated with the article (access type and age)}\\

{\bf Number of cites} & 27.32 & 11 & 61.01 & 0 & 2672 &\\

{\bf OA green} &&&& && 0.30\\
{\bf OA bronze} &&&& && 0.03                        \\
{\bf OA gold}  &&&& && 0.02                         \\
{\bf OA hybrid}&&&& &&  0.10\\
{\bf Closed}&&&& &&  0.55\\
{\bf Year of publication} & 3.55 & 3 & 2.42 & 0 & 7 &\\ \\

\multicolumn{7}{l}{\bf Variables associated with discipline and authors}\\
 {\bf Accounting and Finance}&&&& &&   {0.10}          \\
 {\bf Applied Economics}   &&&& &&     {0.44}           \\
 {\bf Economic Theory}  &&&& &&        {0.03}          \\
 {\bf Business}       &&&& &&          {0.25}           \\
 {\bf Commercial}    &&&& &&           {0.14}          \\
 {\bf Tourism}      &&&& &&            {0.02}         \\
{\bf{Statistics} }     &&&& &&            {0.02}         \\
{\bf Number of authors} & 2.58 &  2 &  1.49 &  1 &  47\\

{\bf Funding}    &&&& &&   0.21 \\ \\

 \multicolumn{7}{l}{\bf Journal prestige and impact variables}\\
 {\bf Expert rating 1}    &&&& &&   0.10         \\
 {\bf Expert rating 2}   &&&& &&   0.30           \\
 {\bf Expert rating 3}   &&&& &&  0.43          \\
 {\bf Expert rating 4}   &&&& &&  0.17          \\
 {\bf Foundation year}   & 1981 & 1985 & 22.01 & 1852 & 2018 &                \\

{\bf SJR} & 1.80 &  1.03 &  2.1283 &  0.103 &  20.643\\

{\bf SJR best quartile} & 1.34 & 1 & 0.59 & 1 & 4 & \\

{\bf Cites per document} & 4.47 &  3.38 &  3.51 &  0.0623 &  30.9186\\ \\

\multicolumn{7}{l}{\bf Variables of influence and social impact}\\
{\bf News mentions} & 0.63 &  0 &  4.89 &  0 &  319\\

{\bf Blog mentions} & 0.18 &  0 &  0.82 &  0 &  37\\

{\bf Policy mentions} & 0.44 &  0 &  2.11339 &  0 &  104\\

{\bf Patent mentions} & 2.5E-3 &  0 &  0.09 &  0 &  12\\

{\bf X mentions} & 8.06 &  1 &  108.72 &  0 &  16317\\

{\bf Facebook} & 0.15 &  0 &  0.75 &  0 &  38\\

{\bf Wikipedia} & 0.07 &  0 &  0.51 &  0 &  30\\

{\bf Video mentions} & 4.6E-3 &  0 &  0.14 &  0 &  25\\

{\bf Mendeley} & 68.85 &  34 &  131.32 &  0 &  8668\\ \\

{\bf Observations} & 43190 &&&&& \\ \hline

\end{tabular}
\end{center}
\end{minipage}}
\end{table}

Regarding the author and disciplinary characteristics, articles have an average of 2.58 authors, with a median of 2 and a range of 1 to 47. Moreover, funding is reported in 21\% of articles. Applied Economics (44\%), Business (25\%), Commerce (14\%), and Accounting and Finance (10\%) are the most common disciplines, while Economic Theory (3\%), Tourism (2\%) and Statistics (2\%) are the least common.

In terms of the prestige and impact of the journal, the year of the journal foundation shows diversity, with a median of 1985 but a standard deviation of 22 years, indicating high variability. The average SJR (Scimago Journal Rank) is 1.8, with a median of 1, showing that most journals have a relatively low score, although there are outliers as high as 20. The median of the best SJR quartile is 1, which means that more than half of the articles analyzed were published in journals classified in the first quartile in one of the different subject categories assigned by the Scopus database. Furthermore, journals in the sample receive an average of 4.47 citations per article three years after publication, with a median of 3.38. Expert ratings vary in frequency, with the highest rating, level 4, occurring 17\% of the time and serving as the base in the regressions. Among the other levels, level 1 (the lowest) is the least frequent at 10\%, followed by level 2 at 30\% and level 3 at 43\%.

Regarding the social impact and influence, mentions vary between sources. The number of news per article has an average of 0.63, with a median of 0 and a range of 319. Similar patterns are observed for the number of mentions in blogs (mean 0.18, median 0, range 37), policy documents (mean 0.44, median 0, range 104), and patents (mean 0.0025, median 0, range 12). The average number of social media mentions in X/Twitter is 8, with a median of 1 and a range of 16,317. Other social media sources are rare (Facebook: average 0.15, median 0, range 38; Wikipedia: average 0.07, median 0, range 30; videos: average 0.0046, median 0, range 25). However, readers of Mendeley, a scientific reference management software, are more frequent (average 68, median 34, range 8668).

{
\subsection{Associations between variables}

Table \ref{tabassociations} shows the Pearson correlation coefficients between the quantitative variables. The strongest positive correlation for the citation count, the dependent variable, is observed with Mendeley readers (0.84). This indicates that papers with a higher number of Mendeley readers tend to have significantly more citations. Policy mentions also show a notable positive correlation (0.45), suggesting that research papers referenced in policy documents are more likely to be cited in academic papers. Blog mentions (0.30) and journal average citations (0.29) are moderately correlated with the citation count, showing that papers in journals with a high average citation rate per document or mentioned in blogs receive more citations. Furthermore, SJR (Scimago Journal Rank) has a positive correlation of 0.26 with the citation count, indicating that papers published in higher-ranked journals are more likely to be cited.

Other variables with notable positive correlations include news mentions (0.21), patent mentions (0.19), journal expert ratings (0.18) and Wikipedia mentions (0.18). Although these correlations are weaker compared to Mendeley readers or policy mentions, they still suggest some influence on citation counts. The year of publication shows a slight negative correlation with citation counts (-0.14), which is expected as older papers have more time to accumulate citations. In addition, the SJR best quartile has a negative correlation of -0.15, which is due to the coding, where quartile 1 corresponds to the top level and quartile 4 to the bottom level.

Several variables show little or no correlation with the citation count. These include funding (0.00), the foundation year of the journal (-0.01), video mentions (0.03), and number of authors (0.06).

Table \ref{tabassociations} also shows several significant associations between the independent variables. The SJR is strongly correlated with the average cites in the journal (0.64), emphasizing that the most prestigious journals tend to publish articles that receive more citations in the three years after their publication. The other highest positive correlations are observed between expert rating and SJR (0.54) and between expert rating and journal average citations (0.41), reflecting that journals rated highly by experts are often highly positioned in the rankings of journals.

Another notable correlation is between news mentions and blog mentions (0.46), indicating that articles mentioned in the news are often discussed in blogs. Similarly, the correlation between blog mentions and policy mentions is notable (0.33), suggesting a link between online discussions and policies.

In addition, Mendeley readers show moderate positive correlations with several variables, including journal average citations (0.39), indicating that articles published in journals with higher citation rates tend to be saved by more readers. The correlation between Mendeley and peer review is weaker (0.16), but still suggests some relationship between peer review and the attention an article receives from the academic community.

The negative association between the best SJR quartile with expert rating and average citations is due to the coding, where quartile 1 corresponds to the top level and quartile 4 to the bottom level, as mentioned above.

In summary, the most influential factors associated with higher citation counts are Mendeley readers, policy mentions, blog mentions, journal average citations, and SJR. These variables should be included in any predictive model for research paper citations. On the other hand, the highest correlations among the independent variables highlight the interplay between peer review, journal rankings, and online mentions.

{

\begin{sidewaystable}
\resizebox{0.65\textwidth}{!}{\begin{minipage}{\textwidth}
\caption{Pearson's correlations}\label{tabassociations}
\begin{center}
\begin{tabular}{lccccccccccccccccccc}
\toprule
Variables & \rotatebox{90}{Number of citations} & \rotatebox{90}{Year of publication} & \rotatebox{90}{Num of authors} & \rotatebox{90}{Funding} & \rotatebox{90}{Expert rating} & \rotatebox{90}{Foundation year} & \rotatebox{90}{SJR} & \rotatebox{90}{SJR best quartile} & \rotatebox{90}{Cites per document} & \rotatebox{90}{News mentions} & \rotatebox{90}{Blog mentions} & \rotatebox{90}{Policy mentions} & \rotatebox{90}{Patent mentions} & \rotatebox{90}{X mentions} & \rotatebox{90}{Facebook} & \rotatebox{90}{Wikipedia} & \rotatebox{90}{Video mentions} & \rotatebox{90}{Mendeley readers} \\
\midrule
Number of citations & 1 & -0.14 & 0.06 & 0.00 & 0.18 & -0.01 & 0.26 & -0.15 & 0.29 & 0.21 & 0.30 & 0.45 & 0.19 & 0.05 & 0.08 & 0.18 & 0.03 & 0.84 \\
Year of publication & -0.14 & 1 & 0.15 & 0.15 & 0.01 & 0.01 & 0.04 & -0.04 & 0.09 & 0.01 & -0.01 & -0.07 & -0.01 & 0.02 & -0.08 & -0.03 & 0.01 & -0.05 \\
Num of authors & 0.06 & 0.15 & 1 & 0.16 & 0.03 & 0.02 & 0.03 & -0.07 & 0.17 & 0.02 & 0.00 & 0.02 & 0.01 & 0.00 & -0.01 & -0.02 & 0.00 & 0.10 \\
Funding & 0.00 & 0.15 & 0.16 & 1 & 0.03 & 0.00 & -0.02 & -0.08 & 0.04 & 0.02 & 0.01 & 0.00 & 0.00 & 0.01 & 0.00 & -0.01 & 0.00 & 0.00 \\
Expert rating & 0.18 & 0.01 & 0.03 & 0.03 & 1 & -0.19 & 0.54 & -0.44 & 0.41 & 0.06 & 0.09 & 0.09 & 0.02 & 0.01 & 0.05 & 0.02 & 0.03 & 0.16 \\
Foundation year & -0.01 & 0.01 & 0.02 & 0.00 & -0.19 & 1 & -0.04 & 0.05 & -0.03 & -0.03 & -0.04 & 0.03 & 0.00 & -0.01 & -0.01 & -0.01 & -0.01 & 0.00 \\
SJR & 0.26 & 0.04 & 0.03 & -0.02 & 0.54 & -0.04 & 1 & -0.37 & 0.64 & 0.09 & 0.20 & 0.23 & 0.02 & 0.04 & 0.06 & 0.06 & 0.05 & 0.26 \\
SJR best quartile & -0.15 & -0.04 & -0.07 & -0.08 & -0.44 & 0.05 & -0.37 & 1 & -0.44 & -0.05 & -0.09 & -0.06 & 0.00 & -0.02 & -0.06 & -0.04 & -0.01 & -0.16 \\
Cites per document & 0.29 & 0.09 & 0.17 & 0.04 & 0.41 & -0.03 & 0.64 & -0.44 & 1 & 0.06 & 0.11 & 0.07 & 0.01 & 0.02 & 0.08 & 0.03 & 0.05 & 0.39 \\
News mentions & 0.21 & 0.01 & 0.02 & 0.02 & 0.06 & -0.03 & 0.09 & -0.05 & 0.06 & 1 & 0.46 & 0.23 & 0.16 & 0.14 & 0.10 & 0.19 & 0.12 & 0.17 \\
Blog mentions & 0.30 & -0.01 & 0.00 & 0.01 & 0.09 & -0.04 & 0.20 & -0.09 & 0.11 & 0.46 & 1 & 0.33 & 0.15 & 0.22 & 0.18 & 0.25 & 0.05 & 0.22 \\
Policy mentions & 0.45 & -0.07 & 0.02 & 0.00 & 0.09 & 0.03 & 0.23 & -0.06 & 0.07 & 0.23 & 0.33 & 1 & 0.07 & 0.05 & 0.06 & 0.14 & 0.01 & 0.29 \\
Patent mentions & 0.19 & -0.01 & 0.01 & 0.00 & 0.02 & 0.00 & 0.02 & 0.00 & 0.01 & 0.16 & 0.15 & 0.07 & 1 & 0.01 & 0.02 & 0.16 & 0.00 & 0.19 \\
X mentions & 0.05 & 0.02 & 0.00 & 0.01 & 0.01 & -0.01 & 0.04 & -0.02 & 0.02 & 0.14 & 0.22 & 0.05 & 0.01 & 1 & 0.24 & 0.27 & 0.01 & 0.05 \\
Facebook & 0.08 & -0.08 & -0.01 & 0.00 & 0.05 & -0.01 & 0.06 & -0.06 & 0.08 & 0.10 & 0.18 & 0.06 & 0.02 & 0.24 & 1 & 0.12 & 0.00 & 0.07 \\
Wikipedia & 0.18 & -0.03 & -0.02 & -0.01 & 0.02 & -0.01 & 0.06 & -0.04 & 0.03 & 0.19 & 0.25 & 0.14 & 0.16 & 0.27 & 0.12 & 1 & 0.01 & 0.16 \\
Video mentions & 0.03 & 0.01 & 0.00 & 0.00 & 0.03 & -0.01 & 0.05 & -0.01 & 0.05 & 0.12 & 0.05 & 0.01 & 0.00 & 0.01 & 0.00 & 0.01 & 1 & 0.03 \\
Mendeley readers & 0.84 & -0.05 & 0.10 & 0.00 & 0.16 & 0.00 & 0.26 & -0.16 & 0.39 & 0.17 & 0.22 & 0.29 & 0.19 & 0.05 & 0.07 & 0.16 & 0.03 & 1 \\
\bottomrule
\end{tabular}
\end{center}
\end{minipage}}
\end{sidewaystable}

}

}

\section{Specific models\label{s3}}

If we allow in \eqref{mp} to $\pi$ to be the gamma distribution with shape parameter $r^{-1}$ and scale parameter $(r\theta)^{-1}$, $r>0$, $\theta>0$ we have that the mixture (unconditional) distribution of the number of citations results
\begin{eqnarray}
\Pr(Y=y|\pmb{x})= \frac{\Gamma(r^{-1}+y)}{\Gamma(r^{-1})\Gamma(y+1)}\left\{\frac{1}{1+r \theta(\pmb{x})}\right\}^{r^{-1}}\left\{\frac{r \theta(\pmb{x})}{1+r \theta(\pmb{x})}\right\}^{y},\quad y=0,1,\dots\label{nbm}
\end{eqnarray}

Here, $r>0$ acts as a dispersion parameter and $\pmb{x}$ a $k\times 1$ vector of exogenous or explanatory variables. Furthermore, $\Gamma(\cdot)$ represents the Euler Gamma function. In this case, the random variable $Y$ has mean and variance given by,
\begin{eqnarray}
\mu_{\pmb{x}} &=& \E(Y|\pmb{x}) = \theta(\pmb{x}),\label{mm}\\
\sigma^2_{\pmb{x}} &=& var (Y|\pmb{x}) = \mu_{\pmb{x}}+r \mu_{\pmb{x}}^2,\label{vm}
\end{eqnarray}
respectively. It is usual to take $\theta(\pmb{x})=\exp(\pmb{x}^{T}\pmb{\beta})$, i.e., we are assuming a log-linear specification in which $\pmb{\beta}$ is a vector of regression parameters which has to be estimated.
This parameterization of the negative binomial regression model has been considered, among others, by \cite{lauless_1987}, \cite{cameronandtrivedi_1998} and \cite{hilbe_2011}.

When $r\to 0$, this model reduces to the Poisson distribution and the geometric model when $r=1$. The log-likelihood function is shown in the Appendix. Details about the normal equations obtained from \eqref{loglike} and second derivatives needed to get the variance-covariance matrix of the estimators can be found in \cite{lauless_1987} and \cite{cameronandtrivedi_1998}.

Although in practice, there are numerous statistical software that have implemented this model in their packages (R, Stata, Eviews, Matlab, and SAS), we have not made use of them but have programmed it in Mathematica, corroborating the results with additional programming in WinRats. In practice, given the difficulty sometimes offered by estimating the index of dispersion, $r$, it is convenient to use the following approximation for the logarithm of the Euler gamma function.

The normal equations obtained from \eqref{loglike} require the use
of the digamma function,
$\psi(z)=\frac{d}{dz}\log(\Gamma(z)),\;z>0$, in order to estimate
all the model parameters. However, this problem is overcome using Mathematica routines \citetext{see \citealp{ruskeepaa_2009}} and RATS
\citetext{see \citealp{brooks_2009}}, which work well with this special
function. Other software, such as Matlab, Stata, Eviews, and R, can also be
useful because they incorporate special packages to work with this model. In practice, given the difficulty that is sometimes offered by estimating the index of dispersion, $r$, it is possible to use the following approximation for the logarithm of the Euler gamma function
\begin{eqnarray*}
\log\left(\Gamma(z)\right)\approx \frac{1}{2}\log(2\pi)+\left(z-\frac{1}{2}\right)
\log(z)-z+\frac{1}{2}z\log\left(z\sinh\left(\frac{1}{z}\right)\right),\quad z>0.
\end{eqnarray*}

Furthermore, the gamma function can also be avoided by taking into account that $\Gamma(a+b)/\Gamma(a)=\prod_{j=1}^{b}(a+j-1)$ and therefore $\log\Gamma(a+b)-\log\Gamma(a)=\sum_{j=1}^{b}\log(a+j-1)$.

\subsection{Distinction between cited and uncited articles}
As was pointed out previously, the process of citations of papers may be divided into two parts. In the first stage, the paper has no citations, and the second part provides the intensity of the citations. Once a paper is cited for the first time, it is contagious to get new citations carried out, at most for the time next to this first citation. Formally, we consider a hurdle model for dealing with papers with citations and documents with no citations in a separate way. We consider a dichotomic variable that first differentiates
documents with and without citations. In the former case, another process generates the number of
citations. A suitable distribution implying the assumption that the data come from two
separate processes is the hurdle count model, and the simplest hurdle model sets the hurdle
at zero. Specifically, the model we consider now is the hurdle model which sets the hurdle at zero with geometric distribution with success probability at zero  $\phi\in[0,1]$ and negative binomial such as in \eqref{nbm} for values larger than zero \citetext{see, for instance, \citealp{mullahy_1986}, and \citealp{pohlmeierandulrich_1995}}. Thus, we will consider the model
\begin{equation}
\Pr(Y=y|\pmb{x})=
\left\{
\begin{array}{lr}
\phi(\pmb{x}),& y=0,\\
 \frac{\bar\phi(\pmb{x})}{\bar p(0|\pmb{x})}\frac{\Gamma(r^{-1}+y)}{\Gamma(r^{-1})\Gamma(y+1)}\left\{\frac{1}{1+r \theta(\pmb{x})}\right\}^{r^{-1}}\left\{\frac{r \theta(\pmb{x})}{1+r \theta(\pmb{x})}\right\}^{y},& y>0,
\end{array}
\right.
\label{hm}
\end{equation}
where $\bar p(0|\pmb{x})=1-p(0|\pmb{x})$, $p(0|\pmb{x})=\Pr(Y=0|\pmb{x})=\left\{1+r \theta(\pmb{x})\right\}^{-r^{-1}}$ (taken from \eqref{nbm}) and $\bar\phi(\pmb{x})=1-\phi(\pmb{x})$. As can be seen, we assume that the hurdle parameter is not constant for all observations
but is modeled similarly to the mean parameter, depending on the covariates. The first (hurdle) part of \eqref{hm} gives the probability of zero citations. The second part governs the process once the hurdle has been passed with a truncated-at-zero probability distribution, which includes the probability of citations conditional on one citation.

The mean and variance of this hurdle distribution are given by
\begin{eqnarray}
\mu_{\pmb{x}} &=& \E(Y|\pmb{x}) = \frac{\bar\phi(\pmb{x})\theta(\pmb{x})}{\bar p(0|\pmb{x})},\label{mim}\\
\sigma^2_{\pmb{x}} &=& var(Y|\pmb{x}) = \mu_{\pmb{x}}\left\{\phi(\pmb{x})+\bar p(0|\pmb{x})+\frac{\mu_{\pmb{x}} }{\bar\phi}\left[r\bar p(0|\pmb{x})+\phi-p(0|\pmb{x})\right]\right\},\label{vim}
\end{eqnarray}
respectively and which are needed to compute, among others, the Pearson residuals.

A logit-link $\phi(\tilde y;r,\pmb{\beta})=\exp(\pmb{x}^{T}\pmb{\delta})/(1+\exp(\pmb{x}^{T}\pmb{\delta}))$, is now assumed to connect the covariates with the parameter $\phi$, where $\pmb{\delta}$ is a new vector of regression parameters to be estimated. Both $\theta$ and $\phi$ may be influenced by different
characteristics and variables. For this reason, the explanatory variables used to model
them may not be the same. Finally, the log-likelihood is shown in the Appendix \citetext{see expression \eqref{loglhm}}. Since the parameters for the two pieces are different (then separable), the maximization may be done separately for each part.

The marginal effect reflects the variation of the conditional mean of citations due to a one-unit change in the $j$th covariate, allowing the mean of citations according to information contained in some explanatory variables. For the log-link, we have that
\begin{eqnarray*}
\frac{\partial \E(y_i)}{\partial x_{ij}}\frac{1}{\E(y_i)}=\beta_j
\end{eqnarray*}
thus, we interpret $\beta_j$ as the proportional change in the mean of citations per unit change in $x_{ij}$. For a dummy variable, taking 0 and 1 values, it is well-known that the estimator $\widehat \exp(\beta_j)$ ($j=1,\dots,k$) is the relative impact of the covariate $j$ on the expected count. This is the same for the logit-link. Nevertheless, for the logit-link and a continuous variable, the effect on the number of citations due to a one-unit change in the covariate is given by
$\partial \phi_i/\partial x_{ij}=\beta_j \bar\phi(1-\bar\phi)$

\section{Results\label{s4}}

\subsection{The homogeneous models}
We begin by fitting the random variable number of citations using Poisson (P), negative binomial (NB), and hurdle negative binomial (HNB) distributions without including covariates. The resulting AIC (Akaike's information criterion) value ($\mbox{AIC}=-2\ell+2p$, where $\ell$ is the loglikelihood value, and $p$ is the number of parameters) was 2297340, 367121, and 364897 for P, NB, and HNB distributions, respectively. For this measure, the smaller the AIC, the better the model is \citetext{see \citealp{akaike_1974}}. The estimated $\lambda$ and $\theta$ parameters for P and NB is the mean of the number of citations, 27.3193, while the estimated index of dispersion $r$ was 1.61933 for the NB distribution and 2.42694 for the HNB distribution. The $\theta$ parameter, in this case, resulted in $23.4883$ and the $\phi$ parameter 0.0552. For all of them, the results were significant. Then, as was expected, the NB distribution is a better fit than the P distribution, and of course, the HNB is better than the NB distribution.

\begin{figure}[htbp]
\begin{center}
\includegraphics[page=1,width=0.45\linewidth]{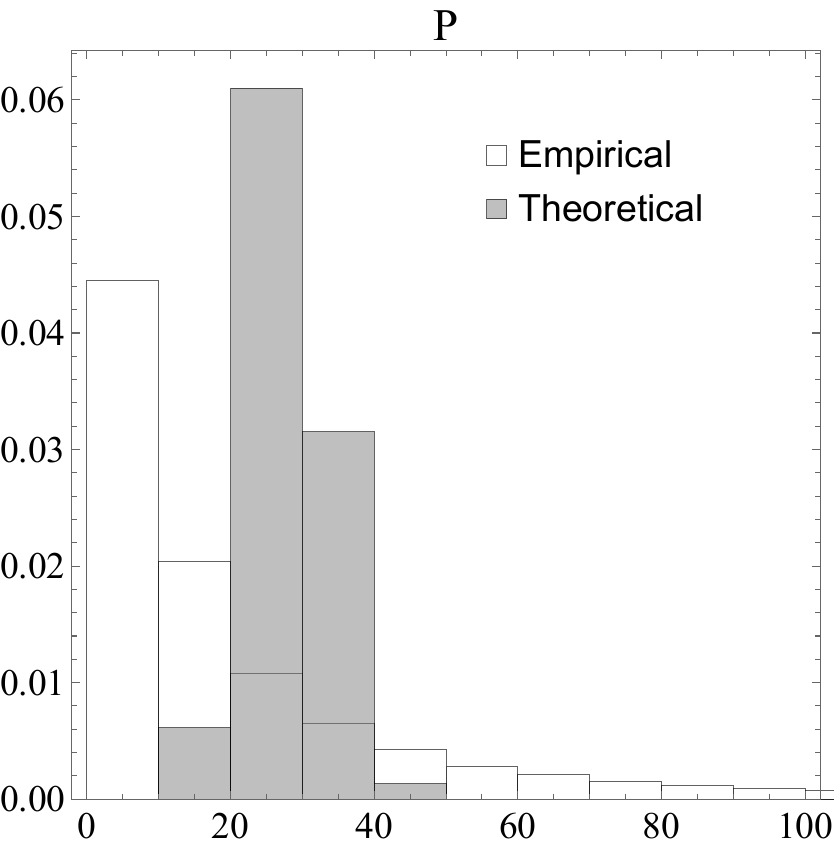}
\hspace{0.25cm}
\includegraphics[page=2,width=0.45\linewidth]{figures.pdf}\\
\includegraphics[page=3,width=0.45\linewidth]{figures.pdf}
\caption{Empirical and fitted histogram of the number of citations obtained by the model based on the Poisson (P), negative binomial (NB) and hurdle negative binomial (HNB) distributions}\label{fighist}
\end{center}
\end{figure}

Figure \ref{fighist} shows the empirical and fitted histograms of the number of citations obtained by the model based on the P, NB, and HNB distributions. From the graph shown in this figure, it is
evident that the fit provided by the NB distribution is much better, especially for the tail of the data and for the zero value, especially when the HNB distribution is considered.

Thus, in advance we will concentrate attention to the negative binomial model.

\subsection{The models with covariates}
Table \ref{tabresults} summarizes the negative binomial regression model for which we can see that many of the coefficients are statistically significant. After this, we have estimated the hurdle model, and the results obtained are shown in this case in Table \ref{tabresultshurdle}. Not all covariates are statistically significant for the hurdle parameter with
different signs in many cases compared to the case in which the dependent variable
takes a value larger than zero. Then, including this model seems to affect the dependent variable studied significantly. For a more parsimonious model, we have removed the variables that are not statistically significant, and the new estimation results are shown in Table \ref{tabresultsrestricted}.

{
Table \ref{tabirr} shows the incidence rate ratios (IRRs), along with their standard errors and confidence intervals, for various categorical variables related to access type, discipline, and journal expert rating. These variables are analyzed to determine their impact on the likelihood of an article being cited. In terms of accessibility, the table shows that most open access modalities have a statistically significant higher incidence rate compared to closed articles, suggesting an open access citation advantage. Concretely, green OA articles have a 13.9\% higher citation rate compared to closed articles, which is statistically significant. Hybrid OA articles have a 7.9\% higher citation rate, also statistically significant. Bronze OA articles see a 5.9\% higher citation rate, again statistically significant. However, gold OA articles have a 6\% higher citation rate, but the confidence interval includes 1, suggesting this result may not be statistically significant.

The table compares the incidence rates for different academic disciplines, using Statistics as the base category. All disciplines are statistically significant, indicating that they are more likely to be cited than Statistics. The disciplines with the highest positive effects are Accounting and Finance with a 42.5\% higher incidence rate, followed by Applied Economics with a 36.8\% higher incidence rate, and Economic Theory with a 36\% higher incidence rate. Tourism and Commerce also have higher incidence rates, with 28.4\% and 28.3\% respectively. Business articles have a 24.9\% higher incidence rate.

The table examines the effect of journal prestige on the incidence rate, using expert ratings from 1 to 4, with 4 being the top tier. The results show that articles in journals with the lowest expert rating have a statistically significant 26.5\% lower incidence rate compared to articles in the highest-rated journals. This suggests that high-impact journals are more selective and publish fewer but more impactful articles, potentially influencing citation practices. Intermediate expert ratings (2 and 3) also show lower incidence rates, but the confidence intervals suggest potential problems with statistical significance.

In summary, Table  \ref{tabirr} highlights the positive impact of OA on the likelihood of an article being cited. Certain disciplines, such as Accounting and Finance, Applied Economics, and Economic Theory, have higher incidence rates, which may reflect their broader impact. The results also suggest that high-impact journals are more selective, publishing fewer but more impactful articles, which may influence citation practices.
}

\begin{table}[htbp]
\resizebox{0.92\textwidth}{!}{\begin{minipage}{\textwidth}
\caption{Negative binomial regression results. They include parameter estimates, standard errors and confidence intervals}\label{tabresults}
\begin{center}
\begin{tabular}{llrlrr}\hline
Category & Variable & Estimate & Std. Err. & 95\% CI lower & 95\% CI upper\\ \hline
 &  \bf{OA green}                            & 0.1299 & 0.0096***   &     0.1109 &  0.1489 \\
{\bf Access type} & \bf{OA bronze}                           & 0.0574 & 0.0248**    &      0.0087 &  0.1061 \\
{\bf and age} & \bf{OA gold}                             & 0.0585 & 0.0306*     &       -0.0015 & 0.1186 \\
 &  \bf{OA hybrid}                           & 0.0749 & 0.0150***   &      0.0454 & 0.1043 \\
 &  \bf{Year of publication}                & -0.1391 & 0.0019***  &   -0.1428 &-0.1353 \\   \\
 &  \bf{Accounting and Finance}             & 0.3544 & 0.0816***   &      0.1945 & 0.5143\\
  & \bf{Applied Economics}                  & 0.3133 & 0.0806***   &      0.1552 & 0.4712\\
 &  \bf{Economic Theory}                    & 0.3075 & 0.0846***   &      0.1417 & 0.4733\\
{\bf Discipline} &  \bf{Business}                           & 0.2227 & 0.0810**    &      0.0640 & 0.3814\\
{\bf and authors}  &  \bf{Commercial}                         & 0.2490 & 0.0812**    &      0.0898 & 0.4080\\
 &  \bf{Tourism}                            & 0.2500 & 0.0859**    &      0.0815 & 0.4184\\
 &  \bf{Number of authors}                  & 0.0452 & 0.0032***   &     -0.3490 & -0.2672 \\
 &  \bf{Funding}                            & 0.0935 & 0.0107***   &     -0.2573 & -0.1903 \\ \\
 &  \bf{Expert rating 1}                    & -0.3081 & 0.0209***  &     -0.1557 & -0.0966 \\
 &  \bf{Expert rating 2}                    & -0.2238 & 0.0171***  &      0.0388 & 0.0515 \\
{\bf Prestige} &  \bf{Expert rating 3}                    & -0.1262 & 0.0151***  &      0.0726 & 0.1145 \\
{\bf and impact} &  \bf{log(Foundation year)}                    & 0.1097 & 0.3920      &     -0.6585 & 0.8780 \\
 &  \bf{SJR}                                & 0.0062 & 0.0034*     &     -0.0004 & 0.0128 \\
 &  \bf{SJR best quartile}                  & -0.2607 & 0.0088***  &     -0.2780 & -0.2433 \\
 &  \bf{Cites per document}                     & 0.0260 & 0.0020***   &      0.0220 & 0.0299 \\ \\
 &  \bf{News mentions}                      & 0.0045 & 0.0011***   &      0.0022 & 0.0067 \\
 &  \bf{Blog mentions}                      & 0.0575 & 0.0069***   &      0.0439 & 0.0709\\
 &  \bf{Policy mentions}                    & 0.0666 & 0.0033***   &      0.0600 & 0.0730 \\
{\bf Influence and} &  \bf{Patent mentions}                    & -0.1267 & 0.0667*    &     -0.2574 & 0.0039 \\
{\bf social impact} &  \bf{X mentions}                         & 0.0002 & 6.7E-5**    &      5.4E-5 & 0.0003 \\
 &  \bf{Facebook}                           & -0.0134 & 0.0058**   &     -0.0249 & -0.0019 \\
 &  \bf{Wikipedia}                          & 0.0237 & 0.0087**    &      0.0067 & 0.0407 \\
 &  \bf{Video mentions}                     & -0.0395 & 0.0257     &     -0.0898 & 0.0108 \\
 &  \bf{Mendeley}                           & 0.0072 & 6.9E-5***   &      0.0070 & 0.0073 \\ \\
 &  \bf{constant}                           & 1.8208 & 2.9721      &     -4.0046 & 7.64623\\
 &  \bf{Index of dispersion,} $\widehat r$  & 0.6833 & 0.0051***   &      0.6732 & 0.6933\\
 &  \bf{AIC} & \multicolumn{4}{r}{327922}\\
 &  \bf{Observations} & \multicolumn{4}{r}{43190}\\ \hline

 & \multicolumn{4}{l}{$^{***}$ indicates 1\% significance level, $^{**}$ indicates 5\% significance level,}\\
 & \multicolumn{4}{l}{$^{*}$ indicates 10\% significance level.}\\
\end{tabular}
\end{center}
\end{minipage}}
\end{table}

\begin{table}[htbp]
\caption{Marginal effect or Incidence Rate Ratio (IRR) for categorical variables}\label{tabirr}
\begin{center}
\begin{tabular}{lrrrr}\hline
Variable  & IRR & Std. Err. & CI lower & CI upper\\
\hline
\multicolumn{2}{c}{\bf Variables associated with the article (access type and age)}\\
 \bf{OA green}                            & 1.1387 & 0.010 & 1.1172 & 1.1605 \\
 \bf{OA bronze}                           & 1.0591 & 0.026 & 1.0087 & 1.1119 \\
 \bf{OA gold}                             & 1.0602 & 0.032 & 0.9985 & 1.1259 \\
 \bf{OA hybrid}                           & 1.0788 & 0.016 & 1.0464 & 1.1099 \\
 \multicolumn{2}{c}{\bf Variables associated with discipline and authors}\\
 \bf{Accounting and Finance}              & 1.4253 &  0.1163 & 1.2147 & 1.6724 \\
 \bf{Applied Economics}                   & 1.3679 &  0.1102 & 1.1678 & 1.6019 \\
 \bf{Economic Theory}                     & 1.3600 &  0.1150 & 1.1522 & 1.6052 \\
 \bf{Business}                            & 1.2494 &  0.1012 & 1.0660 & 1.4643 \\
 \bf{Commercial}                          & 1.2827 &  0.1041 & 1.0939 & 1.5038 \\
 \bf{Tourism}                             & 1.2840 &  0.1102 & 1.0849 & 1.5195 \\
 \multicolumn{2}{c}{\bf Journal prestige and impact variables}\\
 \bf{Expert rating 1}                    & 0.7348 &  0.0153 & 0.8558 & 0.9079 \\
 \bf{Expert rating 2}                    & 0.7995 &  0.0136 & 1.0395 & 1.0528 \\
 \bf{Expert rating 3}                    & 0.8814 &  0.0133 & 1.0753 & 1.1213 \\
 \hline

\end{tabular}
\end{center}
\end{table}

{
The results of the hurdle negative binomial regression model are summarized in Table \ref{tabresultshurdle}. This model is divided into two parts: the positive counts component (negative binomial part) and the zero counts component (hurdle part). The estimates provide insights into which factors are associated with higher or lower citation counts, as well as the likelihood of an article having zero citations.

Younger publications tend to receive fewer citations (estimate = -0.14, $p < 0.01$) and are more likely to have zero citations (0.13, $p < 0.01$). Articles with OA green access are positively associated with citation counts (0.12, $p < 0.01$) and are less likely to have zero citations (-0.28, $p < 0.01$). Conversely, OA bronze access is also positively associated with citation counts (0.13, $p < 0.01$) but increases the likelihood of zero citations (0.72, $p < 0.01$). OA hybrid access is also positively associated with citation counts (0.06, $p < 0.01$).

The number of authors positively influences citation counts (estimate = 0.04, $p < 0.01$) and reduces the likelihood of zero citations (-0.12, $p < 0.01$). Articles with funding also see a positive impact on citation counts (0.08, $p < 0.01$) and a reduced likelihood of zero citations (-0.39, $p < 0.01$). Several disciplines show significant associations with citation counts. Business (-0.27, $p < 0.05$), Tourism (-0.20, $p < 0.05$), Commercial (-0.19, $p < 0.10$), and Applied Economics (-0.16, $p < 0.10$) are negatively associated with citation counts when compared with Statistics.

The journal prestige and impact categories include several variables indicating the journal's reputation and the average article's impact. Lower expert ratings are associated with fewer citations, with rating 1 showing the strongest negative association (estimate = -0.26, $p < 0.01$) while increasing the likelihood of zero citations (0.42, $p < 0.05$). The SJR positively affects citation counts (0.006, $p < 0.10$). The SJR best quartile is negatively associated with citation counts (-0.25, $p < 0.01$) but positively influences the likelihood of zero citations (0.18, $p < 0.01$). Note that a higher quartile represents a worse position in the ranking because the top quartile is coded as 1 while the bottom quartile is coded as 4. The journal average citations positively affects citation counts (0.03, $p < 0.01$) but also positively influences the likelihood of zero citations (0.05, $p < 0.05$).

In the influence and social impact categories, mentions in the news, blogs, policies, patents, and other platforms show mixed effects on citation counts. News mentions positively influence citation counts (estimate = 0.004, $p < 0.01$) and decreases the likelihood of zero citations (-0.15, $p < 0.01$). Blog mentions also positively influence citation counts (0.06, $p < 0.01$). Policy mentions significantly reduce the likelihood of zero citations (-0.96, $p < 0.01$) and positively influence citation counts (0.06, $p < 0.01$). However, patent mentions negatively influence citation counts (-0.11, $p < 0.10$).

Social media mentions have different effects on citation counts. While X/Twitter positively influences citation counts, although with a limited effect (estimate = 0.0001, $p < 0.05$), and decreases the likelihood of zero citations (-0.027, $p < 0.01$), Facebook and video mentions negatively influence citation counts (-0.015 and -0.047 respectively, $p < 0.05$). Wikipedia mentions also positively influence citation counts (0.027, $p < 0.05$). Finally, Mendeley readers have a positive association with citation counts (0.007, $p < 0.01$) and a significant negative association with the likelihood of zero citations (-0.117, $p < 0.01$).

Overall, these results highlight the complex factors that influence the citation patterns of academic articles. Access type, publication age, discipline, collaboration, journal prestige, and social impact metrics all play significant roles in determining the citation outcomes of scholarly work.
}

\begin{table}[htbp]
\resizebox{0.92\textwidth}{!}{\begin{minipage}{\textwidth}
\caption{Parameter estimates from the hurdle negative binomial regression model and standard errors}\label{tabresultshurdle}
\begin{center}
\begin{tabular}{llrlrl}\hline
 & & \multicolumn{2}{c}{Positives (NB part)} & \multicolumn{2}{c}{Zeros (Hurled part)}\\
 \cline{3-6}
Category & Variable & Estimate & Std. Err. & Estimate & Std. Err.\\ \hline
 &  \bf{OA green}                            &   0.1232 & 0.0096*** & -0.2840 & 0.0624***\\
{\bf Access type} & \bf{OA bronze}           &   0.1319 & 0.0258*** &  0.7248 & 0.1003*** \\
{\bf and age} & \bf{OA gold}                 &  -0.0042 & 0.0306 &  0.0744 & 0.1534 \\
 &  \bf{OA hybrid}                           &   0.0589 & 0.0150*** & -0.1256 & 0.0951\\
 &  \bf{Year of publication}                &  -0.1367 & 0.0019*** &  0.1345 & 0.0107***  \\  \\
 &  \bf{Accounting and Finance}             &  -0.1213  & 0.0979  &   0.2095 & 0.3461\\
  & \bf{Applied Economics}                  &  -0.1600  & 0.0971*  &  -0.0344 & 0.3396\\
 &  \bf{Economic Theory}                    &  -0.1197  & 0.1006  &  -0.2191 & 0.3588\\
{\bf Discipline} &  \bf{Business}           &  -0.2663  & 0.0974**  &   0.2287 & 0.3447\\
{\bf and authors}  &  \bf{Commercial}       &  -0.1903  & 0.0976*  &   0.5028 & 0.3431\\
 &  \bf{Tourism}                            &  -0.2000  & 0.1016**  &  -0.3745 & 0.4417\\
 &  \bf{Number of authors}                  &   0.0405  & 0.0031***  &  -0.1235 & 0.0209***\\
 &  \bf{Funding}                            &   0.0782  & 0.0106***  &  -0.3927 & 0.0693***\\ \\
 &  \bf{Expert rating 1}                    &  -0.2634 & 0.0209***   &   0.4217 & 0.1440**\\
 &  \bf{Expert rating 2}                    &  -0.2188 & 0.0169***   &   0.2046 & 0.1321\\
{\bf Prestige} &  \bf{Expert rating 3}      &  -0.1257 & 0.0148***   &  -0.0293 & 0.1251\\
{\bf and impact} &  \bf{log(Foundation year)}    &   0.6218 & 0.3931   &  -0.0907 & 2.1396\\
 &  \bf{SJR}                                &   0.0058 & 0.0033*   &   0.0044 & 0.0364\\
 &  \bf{SJR best quartile}                  &  -0.2464 & 0.0091***   &   0.1801 & 0.0384***\\
 &  \bf{Cites per document}                     &   0.0269 & 0.0019***   &   0.0493 & 0.0177**\\ \\
 &  \bf{News mentions}                      &   0.0040 & 0.0010***   &  -0.1512 & 0.0451***\\
 &  \bf{Blog mentions}                      &   0.0572 & 0.0067***   &  -0.0503 & 0.0863\\
 &  \bf{Policy mentions}                    &   0.0640 & 0.0031***   &  -0.9568 & 0.1251***\\
{\bf Influence and} &  \bf{Patent mentions} &  -0.1140 & 0.0654*   &  -14.406 & 1388.5\\
{\bf social impact} &  \bf{X mentions}      &   0.0001 & 6.1E-5**   &  -0.0270 & 0.0045***\\
 &  \bf{Facebook}                           &  -0.0151 & 0.0057**   &  -0.0056 & 0.0505\\
 &  \bf{Wikipedia}                          &   0.0273 & 0.0086**   &  -0.0085 & 0.0798\\
 &  \bf{Video mentions}                     &  -0.0475 & 0.0234**   &   0.5889 & 0.4109\\
 &  \bf{Mendeley}                           &   0.0068 & 6.6E-5***   &  -0.1175 & 0.0031***\\ \\

 &  \bf{constant}                           & -1.5714   & 2.9816    &  -0.6092 & 16.2303 \\
 &  \bf{Index of dispersion,} $\widehat r$  & 0.6425 & 0.0057***    &    &\\
 &  \bf{AIC} & \multicolumn{4}{r}{324468}\\
 &  \bf{Observations} & \multicolumn{4}{r}{43190}\\ \hline

 & \multicolumn{4}{l}{$^{***}$ indicates 1\% significance level, $^{**}$ indicates 5\% significance level,}\\
 & \multicolumn{4}{l}{$^{*}$ indicates 10\% significance level.}\\
\end{tabular}
\end{center}
\end{minipage}}
\end{table}

\begin{table}[h!]
\resizebox{0.92\textwidth}{!}{\begin{minipage}{\textwidth}
\caption{Parameter estimates from the hurdle negative binomial regression model and standard errors for the restricted model}\label{tabresultsrestricted}
\begin{center}
\begin{tabular}{llrlrl}\hline
 & & \multicolumn{2}{c}{Positives (NB part)} & \multicolumn{2}{c}{Zeros (Hurled part)}\\
 \cline{3-6}
Category & Variable & Estimate & Std. Err. & Estimate & Std. Err.\\ \hline
 &  \bf{OA green}                           &  0.1233 & 0.0096***    & -0.3144 & 0.0609***\\
{\bf Access type} & \bf{OA bronze}          &  0.1329 & 0.0257***    & 0.8389 & 0.0973*** \\
{\bf and age} &  \bf{OA hybrid}             &  0.0593 & 0.0149***    & -- & --\\
 &  \bf{Year of publication}               & -0.1367 & 0.0019***    & 0.1318 & 0.0104***  \\  \\

  & \bf{Applied Economics}                  & -0.0432 & 0.0139**    & -- & --\\
{\bf Discipline} &  \bf{Business}           & -0.1499 & 0.0154***   & -- & --\\
{\bf and authors}  &  \bf{Commercial}       & -0.0754 & 0.0169***   & -- & --\\
 &  \bf{Tourism}                            & -0.0804 & 0.0324**    & -- & --\\
 &  \bf{Number of authors}                  & 0.0406 & 0.0031***   & -0.1125 & 0.0206***\\
 &  \bf{Funding}                            & 0.0786 & 0.0106***   & -0.4125 & 0.0683***\\ \\
 &  \bf{Expert rating 1}                    & -0.2577 & 0.0205***   &  0.3417 & 0.0676***\\
 &  \bf{Expert rating 2}                    & -0.2143 & 0.0166***   & --  & --\\
{\bf Prestige} &  \bf{Expert rating 3}      & -0.1242 & 0.0147***   & --  & --\\
{\bf and impact} &  \bf{SJR}                &  0.0059 & 0.0033*     & --  & --\\
 &  \bf{SJR best quartile}                  & -0.2472 & 0.0091***   &   0.2079 & 0.0356***\\
 &  \bf{Cites per document}                     &  0.0269 & 0.0019***   &   0.0544 & 0.0132***\\ \\
 &  \bf{News mentions}                      & 0.0040 & 0.0010***     &   -0.1472 & 0.0448**\\
 &  \bf{Blog mentions}                      & 0.0567 & 0.0067***    & --  & \\
 &  \bf{Policy mentions}                    & 0.0642 & 0.0032***    &   -0.9866 & 0.1250***\\
{\bf Influence and} &  \bf{Patent mentions} &-0.1120 & 0.0655*      &  -- & \\
{\bf social impact} &  \bf{X mentions}      & 0.0001 & 6.1E-5**      & -0.0255 & 0.0043***\\
 &  \bf{Facebook}                           &-0.0151 & 0.0057**      &  -- & --\\
 &  \bf{Wikipedia}                          & 0.0273 & 0.0086**      & --  & --\\
 &  \bf{Video mentions}                     &-0.0473 & 0.0234**      & --  & --\\
 &  \bf{Mendeley}                           & 0.0068 & 6.6E-5***     & -0.1163 & 0.0031***\\ \\

 &  \bf{constant}                           & 3.0302  & 0.0249***    &   -0.6092 & 16.2303 \\
 &  \bf{Index of dispersion,} $\widehat r$  & 0.6425 & 0.0057***     &    &\\
 &  \bf{AIC} & \multicolumn{4}{r}{324539}\\
 &  \bf{Observations} & \multicolumn{4}{r}{43190}\\ \hline

 & \multicolumn{4}{l}{$^{***}$ indicates 1\% significance level, $^{**}$ indicates 5\% significance level,}\\
 & \multicolumn{4}{l}{$^{*}$ indicates 10\% significance level.}\\
\end{tabular}
\end{center}
\end{minipage}}
\end{table}

\subsection{Models assessment}
Various model fit statistics and residuals are readily available in the statistical literature. In regression study, it is common to examine Pearson residuals, given by $r_i=(y_i-\widehat \mu_{\pmb{x}})/\widehat\sigma_{\pmb{x}}$, where $\mu_{\pmb{x}}$ and $\sigma_{\pmb{x}}$ are given by \eqref{mm} and \eqref{vm} for the negative binomial model and \eqref{mim} and \eqref{vim} for the hurdle model and replacing the regression coefficients by the estimated ones.   These Pearson residuals can also be used to compute the Pearson goodness-of-fit statistic, given by $PS=\sum_{i=1}^{n}r_i^2$, which results in our case 45018.60 and 43523 for the NB and hurdle model, respectively. These value are near to $n-(k+1)=43158$ ($k=31$) and 43128 ($k=61$) for the two estimated models. Here, $k$ is the number of parameters of the model. For the restricted model, the Pearson goodness-of-fit statistic is 43543.80, close to $n-(k+1)=43149$, $k=40$. Then, the model is specified correctly, better than the hurdle model.

Pearson's residuals are often skewed for non-normal data, making the residual plots more challenging to interpret. Therefore, other quantifications
of the discrepancy between observed and fitted values have been suggested in
the literature. In this regard, another choice in the residual analysis is the signed square
root of the contribution to the deviance goodness-of-fit statistic (i.e., deviance residuals).
This is given by \citetext{see \citealp[p.141]{cameronandtrivedi_1998}}
$
d_i=\mbox{sign}(y_i-\widehat \theta_i)\left\{2\left[\ell(y_i)-\ell(\widehat\theta_i)\right]\right\}^{1/2},\quad i=1,\dots,n.
$
Here, sgn is the function that returns the argument's sign (plus or minus). The $\ell(y_i)$
the term is the log-likelihood value when the mean of the conditional distribution for
the $i$-th individual is the individual's actual score of the response variable. The $\ell(\theta_i)$ is
the log-likelihood when the conditional mean is plugged into the log-likelihood. Usually,
the deviance divided by its degree of freedom is examined by taking into account that a
value much greater than one indicates a poorly fitting model. The deviance is given by $D=\sum_{i=1}^{n}d_i$. The statistics $D/df=-0.2891$ for the NB fitted model with covariates indicate a good result. Here, $df$ denotes the degree of freedom. Since the HNB model is not a member of the generalized linear models, it is impossible to compute this statistic. The NB model's expression for deviance is displayed in \eqref{dnb} provided in the Appendix.

\begin{figure}[htbp]
\begin{center}
\includegraphics[page=4,width=0.3\linewidth]{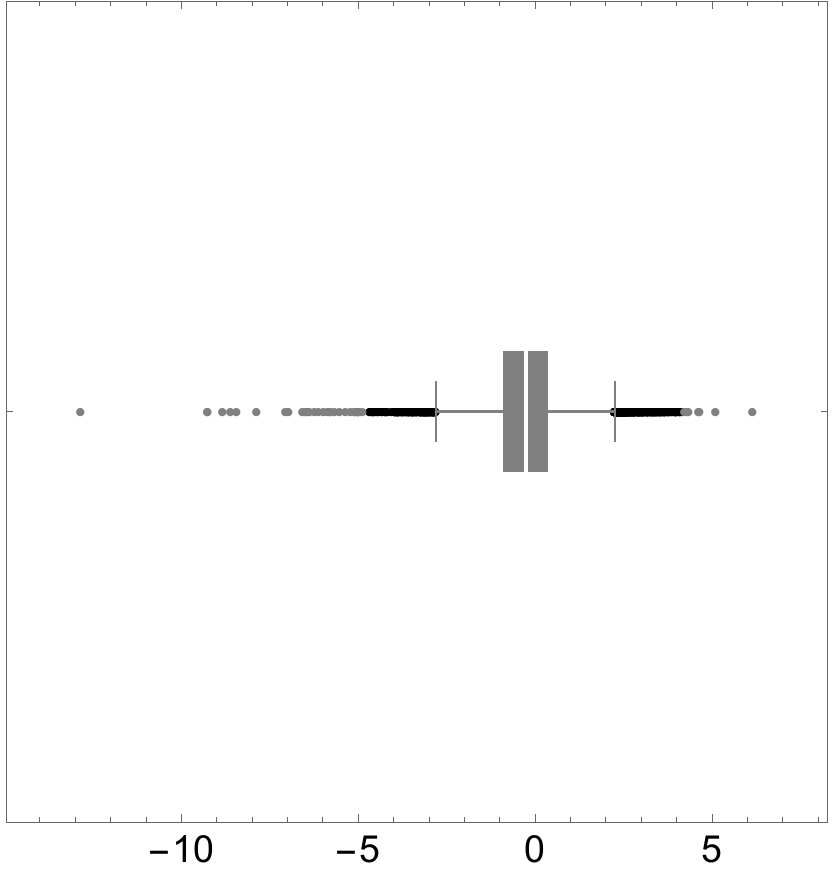}\hspace{0.25cm}
\includegraphics[page=5,width=0.3\linewidth]{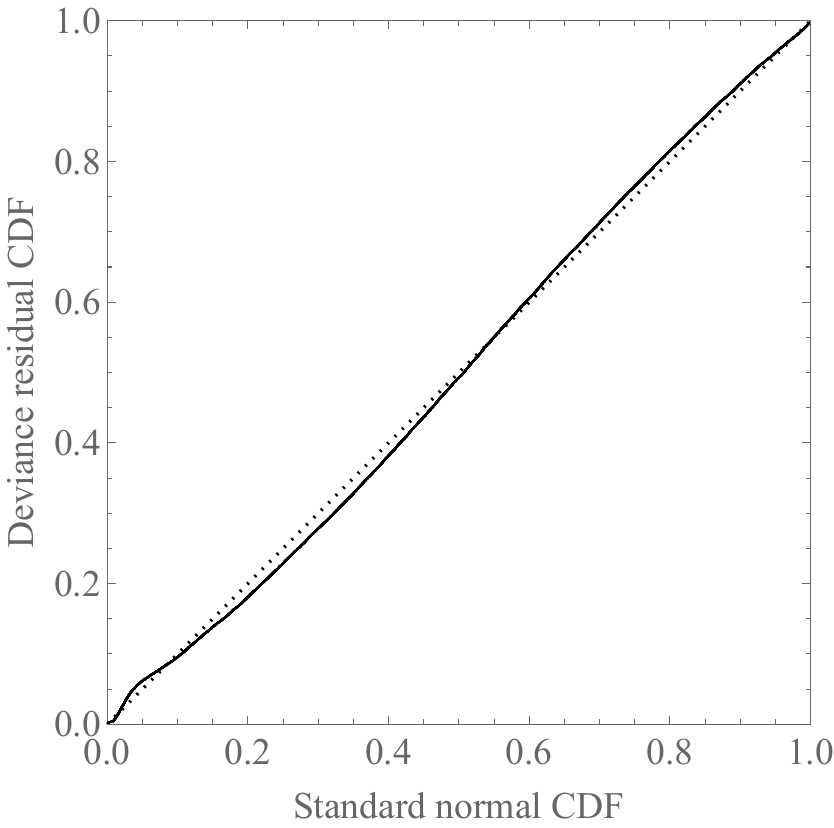}\hspace{0.25cm}
\includegraphics[page=6,width=0.3\linewidth]{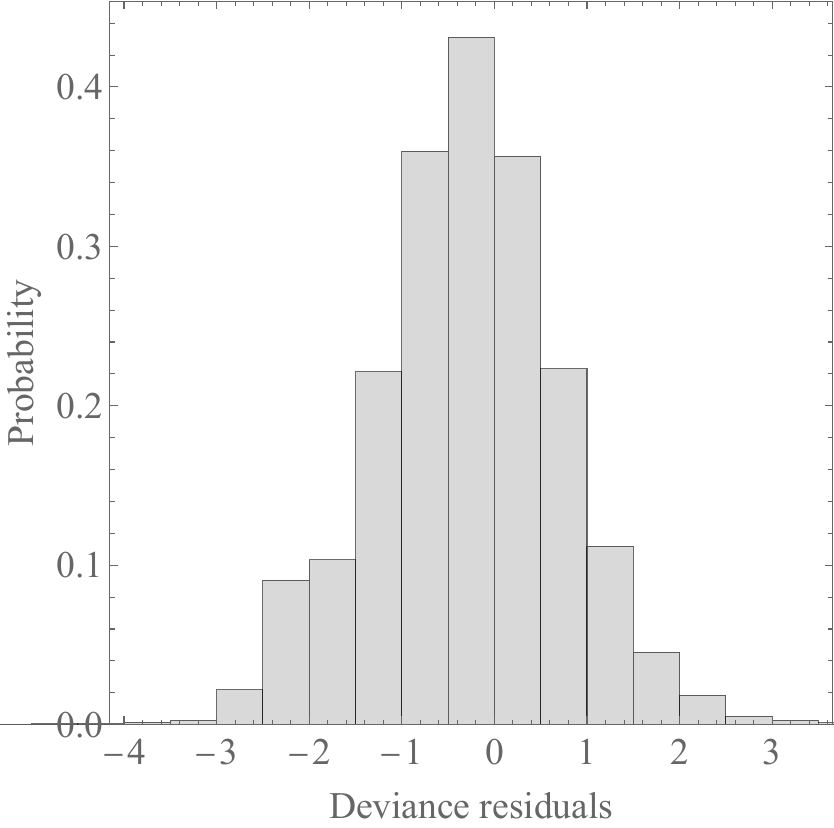}\hspace{0.25cm}
\caption{Box-and-whisker chart of deviance residuals (left panel), probability plot of deviance residuals (center panel) and histogram of the deviance residuals (right panel) based on the NB regression model.}\label{residuals}
\end{center}
\end{figure}

We now illustrate some diagnostic plots based on Pearson's residual and deviance residuals. A box-and-whisker chart of deviance residuals (left panel), probability plot of deviance residuals, and histogram of the deviance residuals based on the NB regression model are shown in Figure \ref{residuals}. All plots indicate a
reasonable behavior of the residuals indicating that the deviance residuals are, to a reasonable approximation, normally distributed, resulting in normal plots that are rather linear.

\begin{figure}[htbp]
\begin{center}
\includegraphics[page=7,width=0.3\linewidth]{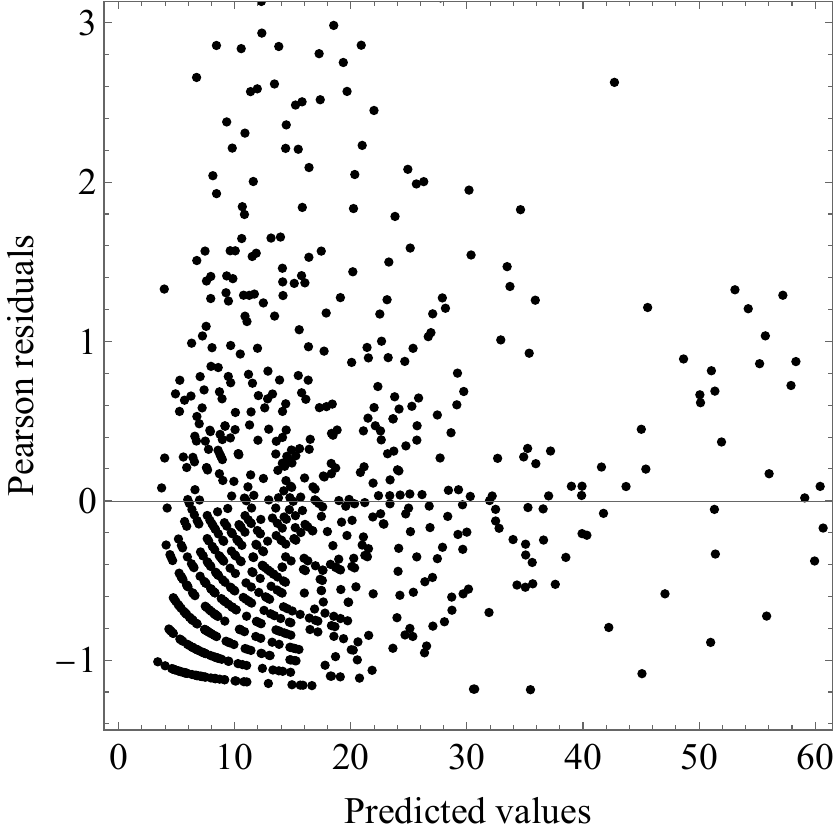}\hspace{0.25cm}\includegraphics[page=8,width=0.3\linewidth]{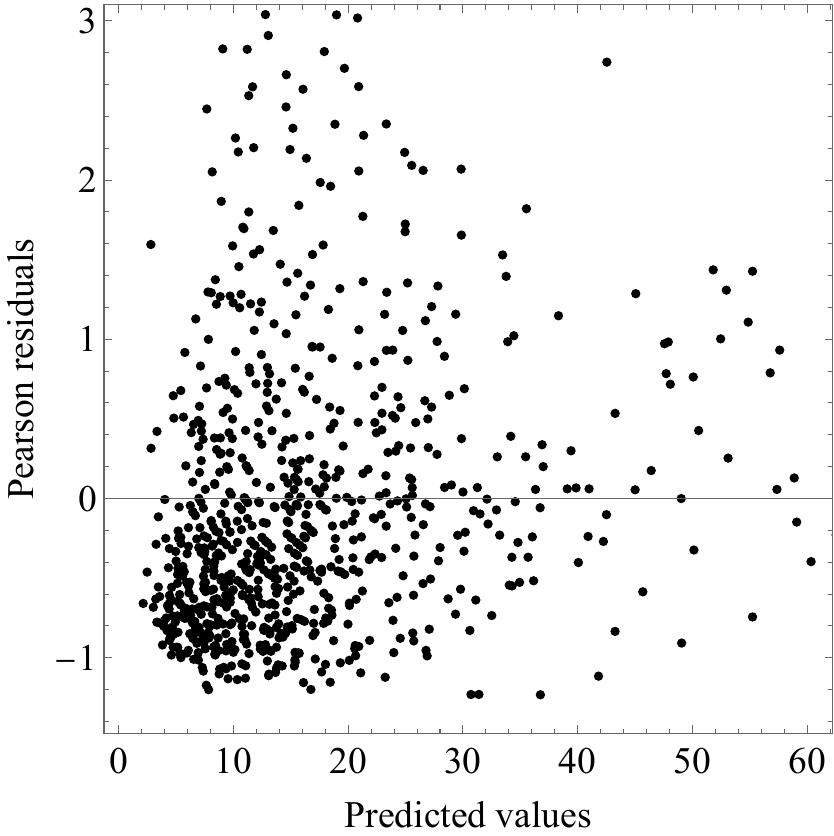}\hspace{0.25cm}\includegraphics[page=9,width=0.3\linewidth]{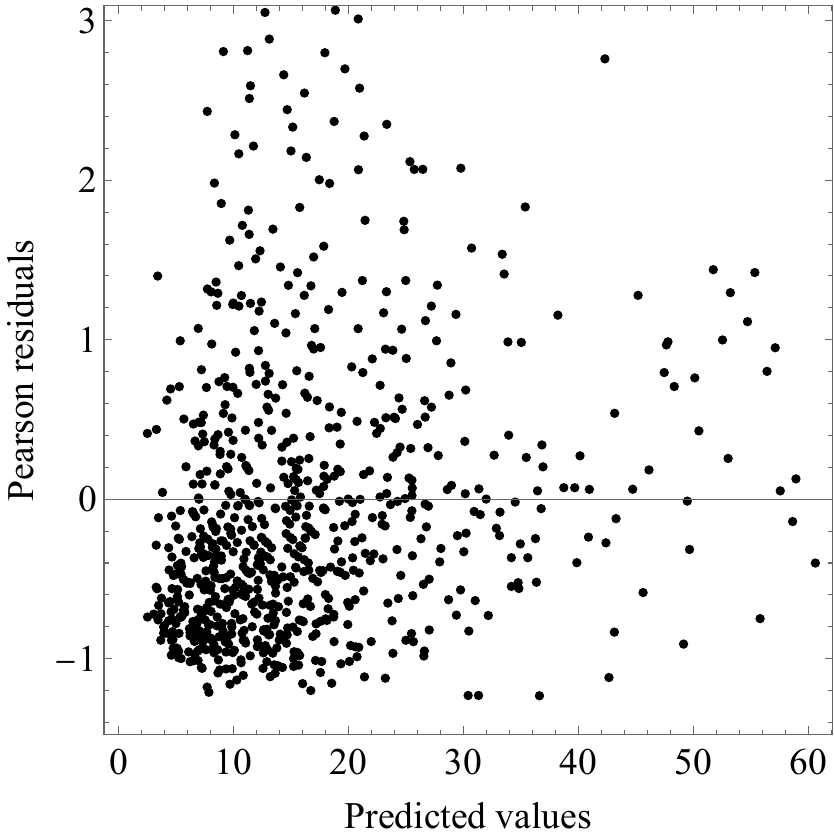}
\caption{Scatter plot of the Pearson's residual against the predicted values: NB (left pannel), HNB (center pannel) and HNB of the restricted model (right pannel) regression models. }\label{presiduals}
\end{center}
\end{figure}

The standardized Pearson's residuals were plotted now against the
predicted values for the NB, HNB, and restricted HNB models, and the results are given in Figure \ref{presiduals}. These plots reveal that no suspicious patterns are apparent.

\section{Conclusions\label{s5}}

{Forecasting future citations and evaluating the quality of an article can be a valuable tool for research assessment, particularly for nations and organizations seeking to gauge their research achievements. Traditional research assessments often rely on citation metrics, expert reviews, and journal-level data. However, recent studies suggest that incorporating a broader range of variables can enhance prediction accuracy. Identifying factors that predict the citation impact and quality of journal articles is crucial for improving scientific research and supporting research evaluation. This study aims to bridge the gap by examining the relationships between document features and extra-documentary factors like altmetric attention scores and downloads to predict citation impact.

Unlike traditional approaches, which rely solely on Poisson or negative binomial models to analyse citation data, our framework explicitly distinguishes between the probability of receiving no citations and the frequency with which citations occur. This enables a more nuanced understanding of the factors influencing the visibility and impact of academic publications. Furthermore, by integrating altmetric indicators and multiple forms of open access, the study sheds new light on how contemporary dissemination practices, beyond conventional journal metrics, shape citation outcomes.

This study aims to understand the factors influencing the citation counts of research papers in Economics and Business using a suitable regression model. The substantial variation in citation counts, where the sample variance exceeds the sample mean, necessitates moving beyond the classical Poisson regression model to a negative binomial regression model. This approach allows the variance to be a function of its predicted value. Furthermore, the citation process can be divided into two stages: papers with no citations and those with citations, to model the observed attraction effect in citation counts after the first citations occur. To address this, we employed a hurdle model, which separately analyzes the likelihood of having zero citations and the intensity of citations for papers that are cited.

The dataset comprises 43,190 research papers from 2014-2021, sourced from The Lens database, with citation counts and social attention scores obtained from the Altmetric database. The results of the hurdle negative binomial regression model reveal that younger publications are associated with fewer citations and a higher likelihood of having zero citations. This is because newer papers have not yet accumulated citations over time. Green OA articles are positively correlated with higher citation counts and a reduced likelihood of zero citations, highlighting a clear citation advantage for articles in open-access repositories. Conversely, bronze OA articles, while also positively associated with citation counts, exhibit an increased likelihood of zero citations, indicating some complexity in their citation dynamics. Hybrid OA articles similarly show a positive association with citation counts.

The number of authors on a paper positively influences citation counts and decreases the likelihood of zero citations, reflecting the collaborative nature of impactful research. Funded articles also see a positive impact on citation counts and are less likely to have zero citations, suggesting that funding may enhance research quality and visibility.

Disciplinary differences are evident, with fields like Business, Tourism, Commercial, and Applied Economics showing lower citation counts compared to Statistics, indicating varying citation practices across disciplines. Journal prestige significantly impacts citation counts, with lower expert ratings linked to fewer citations and a higher likelihood of zero citations. Furthermore, articles in journals with a higher SJR score and better quartile positions tend to have higher citation rates, though the effect varies across quartiles.

Social and influential metrics also play a role in citation counts. News mentions and blog mentions positively influence citation counts and decrease the likelihood of zero citations, underscoring the importance of public and scholarly dissemination. Moreover, policy mentions significantly reduce the likelihood of zero citations and positively influence citation counts, reflecting the impact of policy relevance on academic recognition. However, patent mentions negatively influence citation counts, suggesting a different focus or recognition pattern in patent-referenced work.

Social media exhibit varied effects. While X/Twitter positively influences citation counts and decreases the likelihood of zero citations, Facebook and video mentions negatively affect citation counts. Wikipedia mentions positively influence citation counts, and Mendeley readers are positively associated with citation counts and significantly reduce the likelihood of zero citations, emphasizing the role of academic and public knowledge sources.

Overall, these results highlight the complex factors influencing citation patterns. Access type, publication age, discipline, collaboration, journal prestige, and social impact metrics all significantly determine citation outcomes. The findings advocate for the promotion of open access and collaborative efforts to enhance the visibility and impact of scholarly work. This research provides important knowledge for academics, organizations, and policy makers aiming to understand and improve citation counts in scholarly research. 

However, there are some considerations about the results and their generalization to other fields of research. The field of economics and business has specific characteristics compared to other disciplines, which make generalizations to other fields complex.
 This field faces notable challenges of obsolescence compared to other disciplines, as pointed out by \cite{dortaandgomez_2022}, mainly due to its conservative publication practices and slower adoption of open access. Unlike the rapidly evolving natural and health sciences, where cutting-edge research is widely and rapidly disseminated, economics and business often rely on traditional, slower-moving publication channels. This inertia can delay the adoption of new methodologies and the integration of innovative findings, potentially hindering progress.

Citation concentration within economics and business also varies significantly across subject categories. For example, higher citation concentration indices, as measured by the Gini and Pietra indices, are observed in statistics than in management \citetext{\citealp{gomezanddorta_2024b}}. This disparity suggests that some disciplines receive disproportionately more citations than others. Furthermore, the elasticity, a measure of the sensitivity of impact factors to changes in variables, further underlines the uniqueness of this field. The elasticity of impact factors with respect to the number of citations is higher in economics than in the natural sciences and health sciences, such as biology and medicine. Conversely, the time elasticity of impact factors is lower in economics than in the hard sciences. This means that impact scores are more sensitive to changes in the number of citations but less sensitive to changes over time than those in the sciences and health \citetext{\citealp{gomezanddorta_2024a}}.

These peculiarities suggest that caution should be exercised when attempting to generalise findings from economics and business to other fields. The conservative nature of publication practices, low prevalence of open access, varying citation concentrations, and distinctive elasticity measures highlight the unique landscape of economics and business research. Such considerations are essential for an accurate interpretation of the impact and quality of research in this field.

Our findings reveal systemic biases in citation dynamics that have critical implications for research practices in economics and business. The prominence of journal impact factors (JIFs) and peer-review ratings in predicting citation success may encourage researchers to prioritise conservative methodologies or topics that are considered 'safer' for high-status journals, which could lead to the homogenisation of scholarship. Similarly, the outsized influence of collaboration and funding on citations raises equity concerns, as teams with more resources gain disproportionate visibility, potentially marginalising innovative but underfunded work. The limited impact of altmetric mentions (e.g. in the news or on social media) further highlights how traditional citation metrics undervalue public engagement, thereby discouraging broader societal impact. These dynamics highlight the tension between citation-driven evaluation frameworks and the intellectual diversity that is essential for advancing economic research.

The results of this study have significant implications for researchers, institutions and policymakers in the fields of economics and business. By identifying the key factors influencing the likelihood of being cited and the intensity of citations, the findings can inform strategic decisions regarding research dissemination. For instance, fostering collaboration and securing funding supports research production and enhances its visibility and scholarly impact. Furthermore, the differentiated effects of various open access models suggest that, if they aim to increase citation performance, researchers and institutions should carefully consider their publication and archiving strategies. These insights are particularly relevant in a context where citation metrics continue to play a central role in research evaluation and academic career progression.

In addition to its practical relevance, this study opens up new avenues for future research. The application of a two-stage modelling approach enables a more detailed understanding of citation dynamics, which could be extended to other disciplines or more specific subfields within economics. Furthermore, integrating altmetric data emphasises the increasing importance of examining how new forms of digital visibility intersect with traditional scholarly impact. Future studies could examine the temporal dimensions of citation accumulation, field-specific effects or the role of additional dissemination platforms. Overall, our findings contribute to a broader understanding of how scholarly communication practices are evolving, and of how they shape the recognition of academic work in an increasingly data-driven research environment.

Future research could benefit from a more nuanced examination of author collaboration and its impact on citations. While the present study used the number of authors as an indicator of collaboration, we acknowledge that this measure may oversimplify the complex dynamics of teams. As one of the reviewers noted, large collaborations may involve diminishing returns or free-riding behaviour, and an economic perspective could offer valuable insights into how incentives within research teams influence productivity and scholarly impact. Furthermore, future studies could examine whether the impact of collaboration varies across different subfields of economics and business. While our current dataset lacks the detailed information necessary for investigating these aspects, we recognise the importance of these observations and encourage further research in this area.
}

\paragraph{Funding:} EGD was partially funded by grant PID2021-127989OB-I00 (Ministerio de
Econom\'ia y Competitividad, Spain) and by grant TUR-RETOS2022-075 (Ministerio de Industria, Comercio y Turismo).


\section*{Appendix}
For a sample $\tilde y=(y_1,\dots,y_n)$ of size $n$, the log-likelihood function for the negative binomial regression model is proportional to
\begin{eqnarray}
\ell(\tilde y;r,\pmb{\beta}) & \propto & \sum_{i=1}^{n}\left\{\log\Gamma(r^{-1}+y_i)-(r^{-1}+y_i)\log(1+r\theta_i(\pmb{x}))+y_i\left[\log r+\log\theta_i(\pmb{x})\right]\right\}\nonumber\\
&& -n \log\Gamma(r^{-1}).
\label{loglike}
\end{eqnarray}

For the hurdle model the log-likelihood function is given by,
\begin{eqnarray}
\ell(\tilde y;r,\pmb{\beta},\pmb{\delta}) &=& \sum_{i=1}^{n}\left\{I_{y_i=0}\log(\phi(\pmb{x}))+I_{y_i>0}\log(1-\phi(\pmb{x}))-I_{y_i>0}\log(1-p(0|\pmb{x}))\right\}\nonumber\\
&&+\sum_{i=1}^{n}I_{y_i>0}\left\{\log\Gamma(r^{-1}+y_i)-(r^{-1}+y_i)\log(1+r\theta_i(\pmb{x}))\right.\nonumber\\
&& \left.+y_i\left[\log r+\log\theta_i(\pmb{x})\right]\right\},\label{loglhm}
\end{eqnarray}
where $I$ is the indicator function, $I_A(z)=1$, if $x\in A$ and 0 otherwise.

The NB model's expression for deviance is given by,
\begin{equation}
d_i^2=
\left\{
\begin{array}{lr}
2\left[y_i\log\left(\frac{y_i}{\theta_i}\right)-\left(y_i+r^{-1}\right)\log\left(\frac{1+r y_i}{1+r\theta_i}\right)\right],\quad y_i>0,\\
2 r^{-1}\ln(1+r\theta_i),\quad y_i=0.
\end{array}
\right.
\label{dnb}
\end{equation}


\begin{thebibliography}{}

\bibitem[ABDC, 2022]{abdc_2022}
ABDC (2022).
\newblock Australian Business Deans Council (ABDC) journal quality list.
\newblock Retrieved from \url{https://abdc.edu.au/2022-abdc-journal-quality-list-released/}

\bibitem[Abramo and D'Angelo, 2015]{abramoanddangelo_2015}
Abramo, G., D'Angelo, C. A. (2015).
\newblock The relationship between the number of authors of a publication, its citations and the impact factor of the publishing journal: Evidence from Italy.
\newblock {\em Journal of Informetrics}, 9(4), 746--761.
\newblock \url{https://doi.org/10.1016/j.joi.2015.07.003}

\bibitem[Akaike, 1974]{akaike_1974}
Akaike, H. (1974).
\newblock A new look at the statistical model.
\newblock {\em IEEE Transactions on Automatic Control}, 19(6), 716--723.

\bibitem[Akella et al., 2021]{akella_2021}
Akella, A. P., Alhoori, H., Kondamudi, P. R., Freeman, C., Zhou, H. (2021).
\newblock Early indicators of scientific impact: Predicting citations with altmetrics.
\newblock {\em Journal of Informetrics}, 15(2), 101128.
\newblock \url{https://doi.org/10.1016/j.joi.2020.101128}

\bibitem[Aksnes et al., 2019]{aksnes_2019}
Aksnes, D. W., Langfeldt, L., Wouters, P. (2019).
\newblock Citations, citation indicators, and research quality: An overview of basic concepts and theories.
\newblock {\em SAGE Open}, 9(1), 2158244019829575.
\newblock \url{https://doi.org/10.1177/2158244019829575}

\bibitem[Alpay et al., 2022]{alpay_2022}
Alpay, O., Danacio\u{g}lu, N., \c{C}ankaya, E. (2022).
\newblock Modelling of factors influencing the citation counts in Statistics.
\newblock {\em Academic Platform Journal of Engineering and Smart Systems}, 10(3), 157--167.
\newblock \url{https://doi.org/10.21541/apjess.1075099}

\bibitem[Alperin et al., 2024]{alperin_2024}
Alperin, J. P., Fleerackers, A., Riedlinger, M., Haustein, S. (2024).
\newblock Second-order citations in altmetrics: A case study analyzing the audiences of COVID-19 research in the news and on social media.
\newblock {\em Quantitative Science Studies}, 5(2), 366--382.
\newblock \url{https://doi.org/10.1162/qss_a_00298}

\bibitem[Arbous and Kerrich, 1951]{arbousandkerrich_1951}
Arbous, A. G., Kerrich, J. E. (1951).
\newblock Accident statistics and the concept of accident-proneness.
\newblock {\em Biometrics}, 7(4), 340--432.

\bibitem[Baccini et al., 2014] {baccini_2014}
Baccini, A., Barabesi, L., Cioni, M., Pisani, C. (2014).
\newblock Crossing the hurdle: the determinants of individual scientific performance.
\newblock {\em Scientometrics}, 101, 2035--2062.
\newblock \url{ https://doi.org/10.1007/s11192-014-1395-3}

\bibitem[Bornmann, 2013]{bornmann_2013}
Bornmann, L. (2013).
\newblock What is societal impact of research and how can it be assessed? A literature survey.
\newblock {\em Journal of the American Society for Information Science and Technology}, 64(2), 217--233.
\newblock \url{https://doi.org/10.1002/asi.22803}

\bibitem[Bornmann and Haunschild, 2019]{bornmannandhaunschild_2019}
Bornmann, L., Haunschild, R. (2019).
\newblock Societal impact measurement of research papers.
\newblock In: Gl\"{a}nzel, W., Moed, H. F., Schmoch, U., Thelwall, M. (eds)
\newblock {\em Springer Handbook of Science and Technology Indicators. Springer Handbooks}. Springer, Cham.
\newblock \url{https://doi.org/10.1007/978-3-030-02511-3_23}

\bibitem[Bornmann and Leydesdorff, 2015]{bornmannandleydesdorff_2015}
Bornmann, L., Leydesdorff, L. (2015).
\newblock Does quality and content matter for citedness? A comparison with para-textual factors and over time.
\newblock {\em Journal of Informetrics}, 9(3), 419--429.
\newblock \url{https://doi.org/10.1016/j.joi.2015.03.001}

\bibitem[Bornmann et al., 2012]{bornmann_2012}
Bornmann, L., Schier, H., Marx, W., Daniel, H.-D. (2012).
\newblock What factors determine citation counts of publications in chemistry besides their quality?
\newblock {\em Journal of Informetrics}, 6, 11--18.
\newblock \url{https://doi.org/10.1016/j.joi.2011.08.004}

\bibitem[Breslow, 1984]{breslow_1984}
Breslow, N. (1984).
\newblock Extra-Poisson variation in log-linear models.
\newblock {\em Journal of the Royal Statistical Society. Series C (Applied Statistics)}, 33, 38--44.

\bibitem[Brillinger, 1986]{brillinger_1986}
Brillinger, D. R. (1986).
\newblock The natural variability of vital rates and associated statistics (with Discussion).
\newblock {\em Biometrics}, 42, 693--711.

\bibitem[Brooks, 2009]{brooks_2009}
Brooks, C. (2009).
\newblock {\em RATS Handbook to Accompany
Introductory Econometrics for Finance.}
\newblock Cambridge University Press.

\bibitem[Cameron and Trivedi, 1998]{cameronandtrivedi_1998}
Cameron, A. C., Trivedi, P. K. (1998).
\newblock {\em Regression Analysis of Count Data}
\newblock Econometric Society Monographs No. 30, Cambridge University Press.

\bibitem[Campbell et al., 1991]{campbelletal_1991}
Campbell, M. J., Machin, D., D'Arcangues, C. (1991).
\newblock Coping with extra-Poisson variability in the analysis of
factors influencing vaginal ring expulsions.
\newblock {\em Statistics in Medicine}, 10, 241--251.

\bibitem[Chi and Glänzel, 2017]{chiandglanzel_2017}
Chi, P. S., Gl\"{a}nzel, W. (2017).
\newblock An empirical investigation of the associations among usage, scientific collaboration and citation impact.
\newblock {\em Scientometrics}, 112(1), 403--412.
\newblock \url{https://doi.org/10.1007/s11192-017-2356-4}

\bibitem[Dean et al., 1989]{deanetal_1989}
Dean, C., Lawless, J. F., Willmot, G. E. (1989).
\newblock A mixed Poisson-inverse Gaussian regression model.
\newblock {\em The Canadian Journal of Statistics}, 17(2), 171--181.

\bibitem[de Rijcke et al., 2016]{derijcke_2016}
de Rijcke, S., Wouters, P. F., Rushforth, A. D., Franssen, T. P., Hammarfelt, B. (2016).
\newblock Evaluation practices and effects of indicator use—a literature review.
\newblock {\em Research Evaluation}, 25(2), 161--169.
\newblock \url{https://doi.org/10.1093/reseval/rvv038}

\bibitem[Didegah and Thelwall, 2013] {didegahandthelwall_2013}
Didegah, F., Thelwall, M. (2013).
\newblock Which factors help authors produce the highest impact research? Collaboration, journal and document properties.
\newblock {\em Journal of Informetrics}, 7(4), 861--873.
\newblock \url{ https://doi.org/10.1016/j.joi.2013.08.006}

\bibitem[Dorta-Gonz\'alez and Dorta-Gonz\'alez, 2023a]{dortagonzalez_2023a}
Dorta-Gonz\'alez, P., Dorta-Gonz\'alez, M. I. (2023a).
\newblock Citation differences across research funding and access modalities.
\newblock {\em The Journal of Academic Librarianship}, 49(4), 102734.
\newblock \url{https://doi.org/10.1016/j.acalib.2023.102734}

\bibitem[Dorta-Gonz\'alez and Dorta-Gonz\'alez, 2023b]{dortagonzalez_2023b}
Dorta-Gonz\'alez, P., Dorta-Gonz\'alez, M. I. (2023b).
\newblock The funding effect on citation and social attention: The UN Sustainable Development Goals (SDGs) as a case study.
\newblock {\em Online Information Review}, 47(7), 1358--1376.
\newblock \url{https://doi.org/10.1108/OIR-05-2022-0300}

\bibitem[Dorta-Gonz\'alez and G\'omez-D\'eniz, 2022]{dortaandgomez_2022}
Dorta-Gonz\'alez, P., G\'omez-D\'eniz, E. (2022).
\newblock Modeling the obsolescence of research literature
in disciplinary journals through the age of their cited
references.
\newblock {\em Scientometrics}, 127(6), 2901--2931.
\newblock \url{ https://doi.org/10.1007/s11192-022-04359-w}

\bibitem[Dorta-Gonz\'alez et al., 2017]{dortagonzalez_2017}
Dorta-Gonz\'alez, P., Gonz\'alez-Betancor, S. M., Dorta-Gonz\'alez, M. I. (2017).
\newblock Reconsidering the gold open access citation advantage postulate in a multidisciplinary context: An analysis of the subject categories in the Web of Science database 2009–2014.
\newblock {\em Scientometrics}, 112, 877--901.
\newblock \url{https://doi.org/10.1007/s11192-017-2422-y}

\bibitem[Dorta-Gonz\'alez et al., 2024]{dortagonzalez_2024}
Dorta-Gonz\'alez, P., Rodr\'iguez-Caro, A., Dorta-Gonz\'alez, M. I. (2024).
\newblock Societal and scientific impact of policy research: A large-scale empirical study of some explanatory factors using Altmetric and Overton.
\newblock {\em Journal of Informetrics}, 18(3), 101530.
\newblock \url{https://doi.org/10.1016/j.joi.2024.101530}

\bibitem[Elgendi, 2019]{elgendi_2019}
Elgendi, M. (2019).
\newblock Characteristics of a highly cited article: A machine learning perspective.
\newblock {\em IEEE Access}, 7, 87977--87986.
\newblock \url{https://doi.org/10.1109/ACCESS.2019.2925965}

\bibitem[Engel, 1984]{engel_1984}
Engel, J. (1984).
\newblock Models for response data showing extra-Poisson variation.
\newblock {\em Statistica Neerlandica}, 38, 159--167.

\bibitem[Figg et al., 2006]{figg_2006}
Figg, W. D., Dunn, L., Liewehr, D. J., Steinberg, S. M., Thurman, P. W., Barrett, J. C., Birkinshaw, J. (2006).
\newblock Scientific collaboration results in higher citation rates of published articles.
\newblock {\em Pharmacotherapy}, 26(6), 759--767.
\newblock \url{https://doi.org/10.1592/phco.26.6.759}

\bibitem[G\'omez-D\'eniz and Calder\'in-Ojeda, 2018]{gomezandcalderin_2018}
G\'omez-D\'eniz, E., Calder\'in-Ojeda, E. (2018).
\newblock Properties and applications of the Poisson-reciprocal inverse Gaussian distribution.
\newblock {\em Journal of Statistical Computation and Simulation}, 88(2), 269--289.

{
\bibitem[G\'omez-D\'eniz and Dorta-Gonz\'alez, 2024a]{gomezanddorta_2024a}
G\'omez-D\'eniz, E., Dorta-Gonz\'alez, P. (2024a).
\newblock A field- and time-normalized Bayesian approach to measuring the impact of a publication.
\newblock {\em Scientometrics}, 129, 2659--2676.
\newblock \url{https://doi.org/10.1007/s11192-024-04997-2}

\bibitem[G\'omez-D\'eniz and Dorta-Gonz\'alez, 2024b]{gomezanddorta_2024b}
G\'omez-D\'eniz, E., Dorta-Gonz\'alez, P. (2024b).
\newblock Modeling citation concentration through a mixture of Leimkuhler curves.
\newblock {\em Journal of Informetrics}, 18(2), 101519.
\newblock \url{https://doi.org/10.1016/j.joi.2024.101519}
}


\bibitem[Haustein et al., 2016]{haustein_2016}
Haustein, S., Bowman, T. D., Costas, R. (2016).
\newblock Interpreting ‘altmetrics’: Viewing acts on social media through the lens of citation and social theories.
\newblock In: Sugimoto, C. R. (ed.) {\em Theories of Informetrics and Scholarly Communication} (pp. 372--406). De Gruyter Saur, Berlin.
\newblock \url{https://doi.org/10.1515/9783110308464-022}

\bibitem[Hilbe, 2011]{hilbe_2011}
Hilbe, J. (2011).
\newblock {\em Negative Binomial Regression. Second Edition.}
\newblock  Cambridge University Press, New York.

\bibitem[Hodas and Lerman, 2014] {hodas_2014}
Hodas, N. O., Lerman, K. (2014).
\newblock The simple rules of social contagion.
\newblock {\em Scientific Reports}, 4(1), 4343.
\newblock \url{ https://doi.org/10.1038/srep04343}

\bibitem[Hsu and Huang, 2011]{hsuandhuang_2011}
Hsu, J. W., Huang, D. W. (2011).
\newblock Correlation between impact and collaboration.
\newblock {\em Scientometrics}, 86(2), 317--324.
\newblock \url{https://doi.org/10.1007/s11192-010-0265-x}

\bibitem[Ib\'a\~{n}ez et al., 2013]{ibanez_2013}
Ib\'a\~{n}ez, A., Bielza, C., Larra\~{n}aga, P. (2013).
\newblock Relationship among research collaboration, number of documents and number of citations: A case study in Spanish computer science production in 2000-2009.
\newblock {\em Scientometrics}, 95(2), 689--716.
\newblock \url{https://doi.org/10.1007/s11192-012-0883-6}

\bibitem[Joly et al., 2015]{joly_2015}
Joly, P. -B., Gaunand, A., Colinet, L., Lar\'edo, P., Lemari\'{e}, S., Matt, M. (2015).
\newblock ASIRPA: A comprehensive theory-based approach to assessing the societal impacts of a research organization.
\newblock {\em Research Evaluation}, 24(4), 440--453.
\newblock \url{https://doi.org/10.1093/reseval/rvv015}


\bibitem[Karlis and Xekalaki, 2005]{karlisandxekalaki_2005}
Karlis, D., Xekalaki, E. (2005).
\newblock Mixed Poisson distributions.
\newblock {\em International Statistical Review}, 73, 35--58.

\bibitem[Khazragui and Hudson, 2015]{khazraguiandhudson_2015}
Khazragui, H., Hudson, J. (2015).
\newblock Measuring the benefits of university research: Impact and the REF in the UK.
\newblock {\em Research Evaluation}, 24(1), 51--62.
\newblock \url{https://doi.org/10.1093/reseval/rvu028}

\bibitem[Kousha and Thelwall, 2024]{koushaandthelwall_2024}
Kousha, K., Thelwall, M. (2024).
\newblock Factors associating with or predicting more cited or higher quality journal articles: An Annual Review of Information Science and Technology (ARIST) paper.
\newblock {\em Journal of the Association for Information Science and Technology}, 75(3), 215--244.
\newblock \url{https://doi.org/10.1002/asi.24810}

\bibitem[Kumari et al., 2020]{kumari_2020}
Kumari, R., Uddin, A., Lee, B. H., Choi, K. (2020).
\newblock Analyzing the factors influencing the waiting time to first citation and long-term impact of publications.
\newblock {\em Journal of Scientometric Research}, 9(2), 127--135.
\newblock \url{https://doi.org/10.5530/jscires.9.2.1}

\bibitem[Langfeldt et al., 2020]{langfeldt_2020}
Langfeldt, L., Nedeva, M., S\"{o}rlin, S., Thomas, D. A. (2020).
\newblock Co-existing notions of research quality: A framework to study context-specific understandings of good research.
\newblock {\em Minerva}, 58(1), 115--137.
\newblock \url{https://doi.org/10.1007/s11024-019-09385-2}

\bibitem[Langham-Putrow et al., 2021]{langhamputrow_2021}
Langham-Putrow, A., Bakker, C., Riegelman, A. (2021).
\newblock Is the open access citation advantage real? A systematic review of the citation of open access and subscription-based articles.
\newblock {\em PLoS ONE}, 16(6), e0253129.
\newblock \url{https://doi.org/10.1371/journal.pone.0253129}

\bibitem[Larivière et al., 2015]{lariviere_2015}
Larivière, V., Gingras, Y., Sugimoto, C. R., Tsou, A. (2015).
\newblock Team size matters: Collaboration and scientific impact since 1900.
\newblock {\em Journal of the Association for Information Science and Technology}, 66(7), 1323--1332.
\newblock \url{https://doi.org/10.1002/asi.23266}

\bibitem[Lawless, 1987]{lauless_1987}
Lawless, J.F. (1987).
\newblock Negative binomial and mixed Poisson regression
\newblock {\em The Canadian Journal of Statistics}, 15(3), 209--225.

\bibitem[Leimu and Koricheva, 2005]{leimuandkoricheva_2005}
Leimu, R., Koricheva, J. (2005).
\newblock What determines the citation frequency of ecological papers?
\newblock {\em Trends in Ecology \& Evolution}, 20(1), 28--32.
\newblock \url{https://doi.org/10.1016/j.tree.2004.10.010}

\bibitem[Mammola et al., 2022]{mammola_2022}
Mammola, S., Piano, E., Doretto, A., Caprio, E., Chamberlain, D. (2022).
\newblock Measuring the influence of non-scientific features on citations.
\newblock {\em Scientometrics}, 127(7), 4123--4137.
\newblock \url{https://doi.org/10.1007/s11192-022-04421-7}

\bibitem[Mullahy, 1986]{mullahy_1986}
Mullahy, J. (1986).
\newblock Specification and testing of some modified count data models.
\newblock {\em Journal of Econometrics}, 33, 341--365.

\bibitem[Pohlmeier and Ulrich, 1995]{pohlmeierandulrich_1995}
Pohlmeier, W., Ulrich, V. (1995).
\newblock An econometric model of the two-part decision making process in the demand for
health care.
\newblock {\em The Journal of Human Resources}, 30(2), 339--361.

\bibitem[Ravenscroft et al., 2017]{ravenscroft_2017}
Ravenscroft, J., Liakata, M., Clare, A., Duma, D. (2017).
\newblock Measuring scientific impact beyond academia: An assessment of existing impact metrics and proposed improvements.
\newblock {\em PLoS ONE}, 12(3), e0173152.
\newblock \url{https://doi.org/10.1371/journal.pone.0173152}

\bibitem[Robinson-Garc\'ia et al., 2018]{robinson-garcia_2018}
Robinson-Garc\'ia, N., van Leeuwen, T. N., R\`{a}fols, I. (2018).
\newblock Using altmetrics for contextualised mapping of societal impact: From hits to networks.
\newblock {\em Science and Public Policy}, 45(6), 815--826.
\newblock \url{https://doi.org/10.1093/scipol/scy024}

\bibitem[Roldan-Valadez and Rios, 2015]{roldan-valadez_2015}
Roldan-Valadez, E., Rios, C. (2015).
\newblock Alternative bibliometrics from impact factor improved the esteem of a journal in a 2-year-ahead annual-citation calculation: Multivariate analysis of gastroenterology and hepatology journals.
\newblock {\em European Journal of Gastroenterology \& Hepatology}, 27(2), 115--122.
\newblock \url{https://doi.org/10.1097/MEG.0000000000000253}

\bibitem[Ronda-Pupo, 2017]{ronda-pupo_2017}
Ronda-Pupo, G. A. (2017).
\newblock The effect of document types and sizes on the scaling relationship between citations and co-authorship patterns in management journals.
\newblock {\em Scientometrics}, 110(3), 1191--1207.
\newblock \url{https://doi.org/10.1007/s11192-016-2231-8}

\bibitem[Rousseau, 1992]{rousseau_1992}
Rousseau, R. (1992).
\newblock Why am I not cited or, why are multi-authored papers more cited than others?
\newblock {\em Journal of Documentation}, 48(1), 79--80.

\bibitem[Ruan et al., 2020]{ruan_2020}
Ruan, X., Zhu, Y., Li, J., Cheng, Y. (2020).
\newblock Predicting the citation counts of individual papers via a BP neural network.
\newblock {\em Journal of Informetrics}, 14(3), 101039.
\newblock \url{https://doi.org/10.1016/j.joi.2020.101039}

\bibitem[Ruskeepaa, 2009]{ruskeepaa_2009}
Ruskeepaa, H. (2009).
\newblock {\em Mathematica Navigator. Mathematics, Statistics, and Graphics. Third Edition}.
\newblock Academic Press. USA.

\bibitem[Shen et al., 2021]{shen_2021}
Shen, H., Xie, J., Li, J., Cheng, Y. (2021).
\newblock The correlation between scientific collaboration and citation count at the paper level: A meta-analysis.
\newblock {\em Scientometrics}, 126(4), 3443--3470.
\newblock \url{https://doi.org/10.1007/s11192-021-03888-0}

\bibitem[Sin, 2011]{sin_2011}
Sin, S. C. J. (2011).
\newblock International coauthorship and citation impact: A bibliometric study of six LIS journals, 1980--2008.
\newblock {\em Journal of the American Society for Information Science and Technology}, 62(9), 1770--1783.
\newblock \url{https://doi.org/10.1002/asi.21572}

\bibitem[Sj\"{o}g{\aa}rde and Didegah, 2022]{sjogarde_2022}
Sj\"{o}g{\aa}rde, P., Didegah, F. (2022).
\newblock The association between topic growth and citation impact of research publications.
\newblock {\em Scientometrics}, 127(4), 1903--1921.
\newblock \url{https://doi.org/10.1007/s11192-022-04293-x}

\bibitem[Spaapen and Van Drooge, 2011]{spaapen_2011}
Spaapen, J., Van Drooge, L. (2011).
\newblock Introducing ‘productive interactions’ in social impact assessment.
\newblock {\em Research Evaluation}, 20(3), 211--218.
\newblock \url{https://doi.org/10.3152/095820211X12941371876742}

\bibitem[Thelwall et al., 2023]{thelwall_2023}
Thelwall, M., Kousha, K., Abdoli, M., Stuart, E., Makita, M., Font-Julián, C., Wilson, P., Levitt, J. (2023).
\newblock Is research funding always beneficial? A cross-disciplinary analysis of UK research 2014-20.
\newblock {\em Quantitative Science Studies}, 1--34.
\newblock \url{https://doi.org/10.1162/qss_a_00254}

\bibitem[Thelwall and Maflahi, 2020]{thelwall_2020}
Thelwall, M., Maflahi, N. (2020).
\newblock Academic collaboration rates and citation associations vary substantially between countries and fields.
\newblock {\em Journal of the Association for Information Science and Technology}, 71(8), 968--978.
\newblock \url{https://doi.org/10.1002/asi.24315}

\bibitem[Urlings et al., 2021]{urlings_2021}
Urlings, M. J., Duyx, B., Swaen, G. M., Bouter, L. M., Zeegers, M. P. (2021).
\newblock Citation bias and other determinants of citation in biomedical research: Findings from six citation networks.
\newblock {\em Journal of Clinical Epidemiology}, 132, 71--78.
\newblock \url{https://doi.org/10.1016/j.jclinepi.2020.11.019}

\bibitem[Vieira and Gomes, 2010]{vieira_2010}
Vieira, E. S., Gomes, J. A. N. F. (2010).
\newblock Citation to scientific articles: Its distribution and dependence on the article features.
\newblock {\em Journal of Informetrics}, 4(1), 1--13.
\newblock \url{https://doi.org/10.1016/j.joi.2009.06.002}
\bibitem[Wagner et al., 2019]{wagner_2019}
Wagner, C. S., Whetsell, T. A., Mukherjee, S. (2019).
\newblock International research collaboration: Novelty, conventionality, and atypicality in knowledge recombination.
\newblock {\em Research Policy}, 48(5), 1260--1270.
\newblock \url{https://doi.org/10.1016/j.respol.2019.01.002}

\bibitem[Wilsdon et al., 2015]{wilsdon_2015}
Wilsdon, J., Allen, L., Belfiore, E., Campbell, P., Curry, S., Hill, S., et al. (2015).
\newblock The metric Tide: Report of the independent review of the role of metrics in research assessment and management.
\newblock Bristol: Higher Education Funding Council for England (HEFCE).
\newblock \url{https://doi.org/10.13140/RG.2.1.4929.1363}

\bibitem[Willmot, 1987]{willmot_1987}
Willmot, G. (1987).
\newblock The Poisson-inverse Gaussian distribution as an alternative to the negative binomial.
\newblock {\em Scandinavian Actuarial Journal}, 3-4, 113--127.

\bibitem[Winkelmann, 2008]{winkelmann_2008}
Winkelmann, R. (2008).
\newblock {\em Econometric Analysis of Count Data. 5th edition.}
\newblock  Heidelberg, Ger: Springer.

\bibitem[Wouters et al., 2019]{wouters_2019}
Wouters, P., Zahedi, Z., Costas, R. (2019).
\newblock Social media metrics for new research evaluation.
\newblock In: Gl\"{a}nzel, W., Moed, H.F., Schmoch, U., Thelwall, M. (eds), {\em Springer Handbook of Science and Technology Indicators}. Springer Handbooks. Springer, Cham.
\newblock \url{https://doi.org/10.1007/978-3-030-02511-3_26}


\end{thebibliography}
\end{document}